\documentclass[12pt,preprint,a4paper]{aastex}
\usepackage{times}
\usepackage{graphics,epsf}
\usepackage{amsmath}                
\usepackage{amsfonts}               
\usepackage{amssymb}                
\usepackage{epsfig}                 
\usepackage{graphicx}
\usepackage{url}
\usepackage{tabularx}
\usepackage{booktabs}
\usepackage{epstopdf}
\usepackage{subfigure}

\def \cm{~\rm{cm}}
\def \s{~\rm{s}}
\def \km{~\rm{km}}

\def \K{~\rm{K}}
\def \g{~\rm{g}}

\def \AU{~\rm{AU}}

\def \yr{~\rm{yr}}

\begin{document}

\title{Shaping planetary nebulae with jets in inclined triple stellar systems}

\author{Muhammad Akashi\altaffilmark{1} and  Noam Soker\altaffilmark{1}}

\altaffiltext{1}{Department of Physics, Technion -- Israel
Institute of Technology, Haifa 32000, Israel; akashi@ph.technion.ac.il,
soker@physics.technion.ac.il}

\begin{abstract}
We conduct three-dimensional hydrodynamical simulations of two opposite jets launched obliquely to the orbital plane around an asymptotic giant branch (AGB) star and within its dense wind, and demonstrate the formation of a `messy' planetary nebula (PN), namely, a PN lacking any type of symmetry (highly irregular).
In building the initial conditions we assume that a tight binary system orbits the AGB star, and that
the orbital plane of the tight binary system is inclined to the orbital plane of the binary system and the AGB star (the triple system plane). We further assume that the accreted mass on to the tight binary system forms an accretion disk around one of the stars, and that the plane of the disk is tilted to the orbital plane of the triple system. The highly asymmetrical and filamentary structures that we obtain support the notion that messy PNe might be shaped by triple stellar systems.
\end{abstract}

\keywords{binaries: close $-$ planetary nebulae $-$ jets}

\section{INTRODUCTION}
\label{sec:intro}

The notion that binary stellar systems (e.g, \citealt{BondLivio1990, Bond2000, DeMarco2015, Zijlstra2015}) or planetary systems (e.g, \citealt{DeMarcoSoker2011}) shape planetary nebulae (PNe) has gained a huge support over the years, both from theoretical considerations that point to the difficulties of models based on single asymptotic giant branch (AGB) stellar progenitors to shape PNe (e.g., \citealt{SokerHarpaz1992, NordhausBlackman2006, GarciaSeguraetal2014}), and from a large number of observations and their analysis
(e.g., from 2015 on, \citealt{Akrasetal2015, Alleretal2015a, Alleretal2015b, Boffin2015, Corradietal2015, Decinetal2015, DeMarcoetal2015, Douchinetal2015, Fangetal2015, Gorlovaetal2015, Hillwigetal2015, Jones2015, Jonesetal2015, Manicketal2015, Martinezetal2015, Miszalskietal2015, Mocniketal2015, Montezetal2015, Jonesetal2016, Chiotellisetal2016, Akrasetal2016, GarciaRojasetal2016, Jones2016, Hillwigetal2016a, Hillwigetal2016b, Bondetal2016, Chenetal2016, Madappattetal2016, Alietal2016, Hillwigetal2017}).
In most cases of shaping by binary interaction, jets are involved (e.g., \citealt{Morris1987, Soker1990AJ, SahaiTrauger1998, Boffinetal2012, HuarteEspinosaetal2012, Balicketal2013, Miszalskietal2013, Tocknelletal2014, Huangetal2016, Sahaietal2016, RechyGarciaetal2016, GarciaSeguraetal2016}, to list a small number out of hundreds of papers).
A relatively small number of papers raise the possibility that triple stellar systems are behind the morphologies of some PNe (e.g., \citealt{Sokeretal1992, Soker1994, Soker2016triple, BearSoker2017, Jones2017}). A claim for a triple stellar system in the PN SuWt~2 \citep{Bondetal2002, Exteretal2010} was rejected recently by \cite{JonesBoffin2017}. 
   
In general there are two orbital planes. One is that of the more tight binary system, and the second one is that of the triple stellar system, i.e., of the tight binary system motion around the center of mass with the third star. If the two orbital planes coincides, the interaction leads to a mass loss geometry that has a plane of symmetry, even if the mass loss geometry departs from axi-symmetry. Binary systems, e.g., if the interaction time scale is shorter than the orbital period or if the orbit is eccentric, can also cause departure from axisymmetry (e.g., \citealt{SokerHadar2002}).

If the two orbital planes of the triple system are inclined to each other, the morphology of the  descendant PN is likely to possess no symmetry plane. The descendant PN is actually likely to lack any symmetry; nor point-symmetry, nor axial-symmetry, and nor mirror symmetry.
We term this a `messy PN'.

As part of the increasing interest in mass transfer in evolving triple stellar systems (e.g., \citealt{MichaelyPerets2014, Portegies2016}), we extend our study of the shaping of PNe by jets, now with properties as expected in some triple stellar systems.
Specifically, we consider the flow structure described by \cite{Soker2004}, where a tight binary system orbits an AGB star and accretes mass from the AGB wind. Because the orbital plane of the tight binary system is not parallel to the orbital plane around the AGB star, the jets' axis is not perpendicular to the orbital plane of the triple system. We describe the initial setting and the numerical code in section \ref{sec:numerical}. We then describe the numerical results in section \ref{sec:results}.
We summarize our results in section \ref{sec:summary}. 
  
\section{NUMERICAL SET-UP}
 \label{sec:numerical}
\subsection{Initial conditions}
 \label{sec:initial}
\cite{Soker2004} studied triple stellar systems where a tight binary system orbits
an AGB star that blows a dense wind. The two orbital planes, that of the tight
binary system and that of the triple stellar system are inclined to each other.
The tight binary system accretes mass from the wind. Due to the orbital motion
of the two stars of the tight binary system around their mutual center of mass, the
accreted mass possess specific angular momentum that leads to the formation of an
accretion disk around one (or two) of the stars. The plane of the accretion disk
is close to being parallel to the orbital plane of the tight binary system, rather than to that of the triple stellar system \citep{Soker2004}. The accretion disk launches two opposite jets that are almost perpendicular to the orbital plane of the tight binary system.

\begin{figure}[h!]
\begin{center}
\includegraphics[width=120mm]{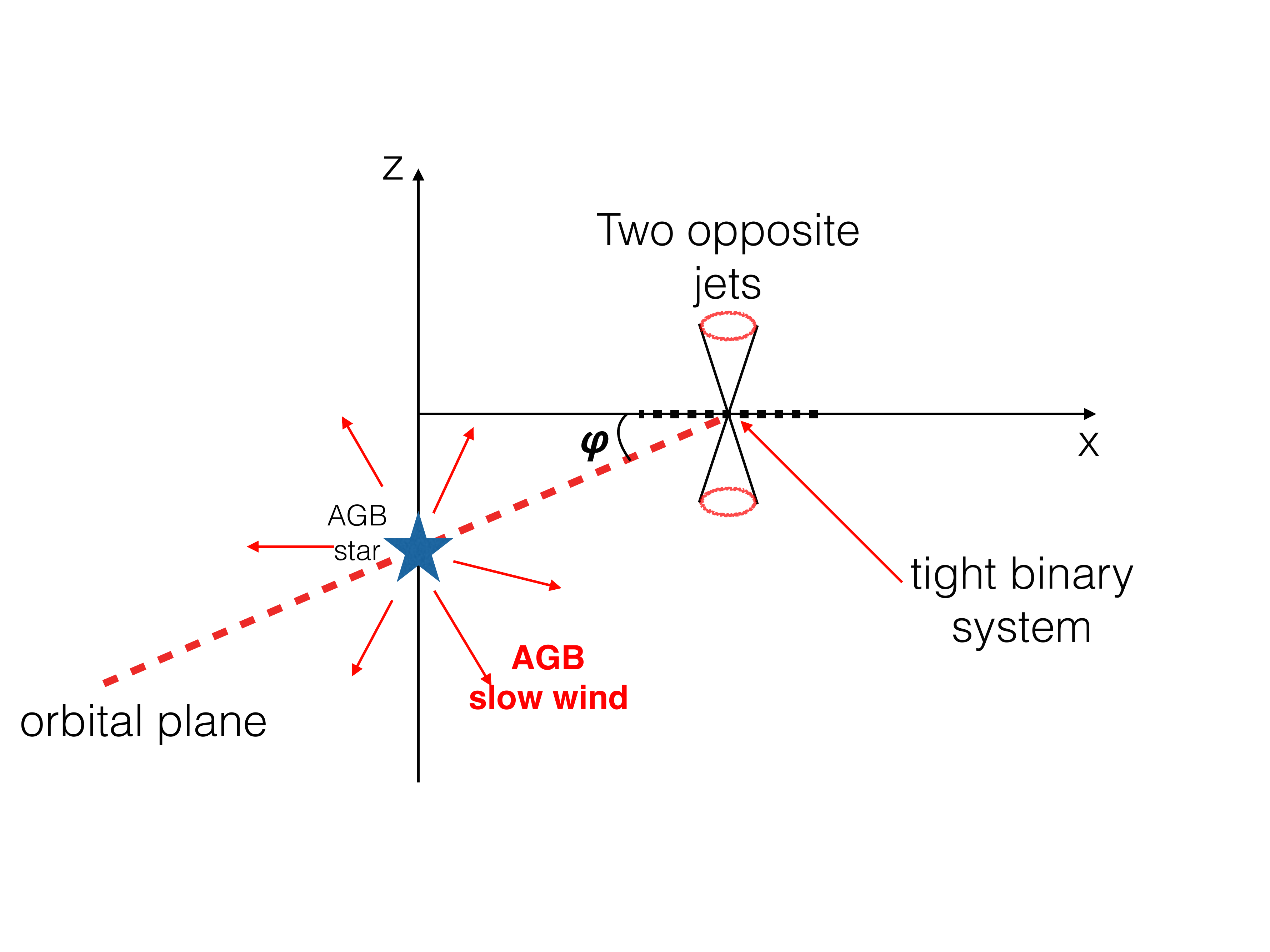}
\caption{A schematic drawing of the outflow from the triple stellar system in the $y=0$ plane.
The orbital plane of the tight binary system and the one of the triple system are inclined by an angle $\phi$. The projection of the orbit on the $y=0$ plane is marked with the red dashed-line. 
The tight binary system accretes mass from the AGB wind and launches jets as indicated. 
The AGB star is located at $(x,y,z)=(0,0,-a_{123} \sin \phi)$ during the entire simulation. At $t=0$ the tight binary system starts its orbit at $(x,y,z)=(a_{123} \cos \phi, 0,0$), as marked in this figure. }
 \label{fig:schematic}
\end{center}
%
%
\end{figure}

We do not study the mass accretion process, but assume that two opposite jets
are launched from an origin that orbits the AGB star.
We do note though, that the accretion rate might be larger by a factor of several relative to the classical Bondi-Hoyle-Lyttleton accretion rate, as the mass transfer might take place via the so called wind Roche-lobe overflow (RLOF; \citealt{MohamedPodsiadlowski2011}). In a full wind RLOF \cite{MohamedPodsiadlowski2011} claim that the accretion rates are at least an order of magnitude greater than the classical Bondi-Hoyle-Lyttleton accretion rates.
In Fig. \ref{fig:schematic} we
present a schematic drawing of the wind blown by the AGB star and the jets launched by the
tight binary system, and the orbit of the jets' origin around the AGB star.

We simulate two cases differ in the inclination angle $\phi$ of the jets' axis to the line perpendicular to the orbital plane of the triple system. In the first case $\phi=20^\circ$ and in the second case $\phi=30^\circ$.
The other parameters are identical for the two runs and are as follows.
The orbital separation of the jets' origin (the location of the tight binary system) to the center of the AGB star is $a_{123}=30 \AU$. The orbital period is $P_{123}=67~$yrs, corresponding to a total mass of the triple stellar system of $M_{123}=M_1+M_2+M_3=5 M_\odot$.

At the beginning of each simulation the slow AGB winds fills the grid, and it is continuously replenished along the simulation by injecting mass from the numerical surface of the AGB star at a radius of $r= 1145 R_\odot$ from the center. At this radius the wind almost reaches its terminal velocity.
The mass loss rate and velocity into the slow AGB wind are $\dot M_w = 5 \times 10^{-5} M_\odot \yr^{-1}$ and $v_w=10 \km \s^{-1}$, respectively.

At $t=0$ we start to launch two jets,  along the $+z$ and $-z$ directions. The jets are conical, with a half opening angle of $\alpha_j=30^\circ$. The opening angle is similar to the values we have been using over the years ($\simeq 30-50^\circ$) while studying the shaping of PNe by jets (e.g., \citealt{AkashiSoker2013}).
The jets opening angle influences the morphology of the nebula. However,  we have limited numerical resources and cannot study more cases at this time. The initial velocity of the jets is $v_j = 500 \km \s^{-1}$, that is about equal to the escape speed from a main sequence star. The mass loss rate into the two jets together is $\dot M_{2j}=2 \times 10^{-7} M_\odot = 0.004 \dot M_w$. This corresponds to about $10 \%$ of the Bondi-Hoyle-Lyttleton accretion rate from the wind.
   
\subsection{The numerical code}
 \label{sec:code}

Our simulations are performed by using version 4.0-beta of the
FLASH code \citep{Fryxell2000}. The FLASH code is an adaptive-mesh
refinement (AMR) modular code used for solving hydrodynamics or
magnetohydrodynamics problems. Here we use the unsplit PPM
(piecewise-parabolic method) solver of FLASH. We neither include
gravity, as velocities are much above the escape speed in the
region we simulate, nor radiative cooling. Radiative cooling is important.
At this first study we did not include radiation from the mass-accreting star(s) as well. As we will see below, the gas does suffer a substantial adiabatic cooling. Radiative cooling will further reduce the gas temperature, while radiation from the stars can heat it. As we did not include radiative cooling and radiation, our results cannot reproduce exact details of observed PNe, but non the less give the overall morphological characteristics of `messy nebulae'.

We employ a full 3D AMR (8 levels; $2^{11}$ cells in each
direction, or 10 levels; $2^{13}$) using a Cartesian grid $(x,y,z)$ with outflow boundary
conditions at all boundary surfaces. The center of the grid is at $(0,0,0)$.
The AGB star is placed at $(0,0,-a_{123} \sin \phi $). The tight binary system orbits the AGB star along a circle. The projection of the orbit on the $y=0$ plane is shown in Fig.  \ref{fig:schematic}.
We simulate the whole space with no symmetry assumptions.
The jets are continuously launched during the entire simulated time.

The slow wind and the jets start with a
temperature of $1000 \K$. The initial jets' temperature has no
influence on the results (as long as it is highly supersonic) because
the jets rapidly cool due to adiabatic expansion.
For numerical reasons a weak slow wind is injected from the jets' origin in the sector
$\alpha_j<\theta<90^\circ$.  This wind has a small effect on the outcome, as
it contains very little energy. Without this wind numerical instabilities might develop on the boundary between the ambient gas and the sphere from where the jets are launched (more numerical details are in \citealt{AkashiSoker2013} ).
  
\section{RESULTS}
 \label{sec:results}
\begin{figure}
\begin{center}
\subfigure[$t=41$~\yr]{\includegraphics[height=2.7in,width=2.7in,angle=0]{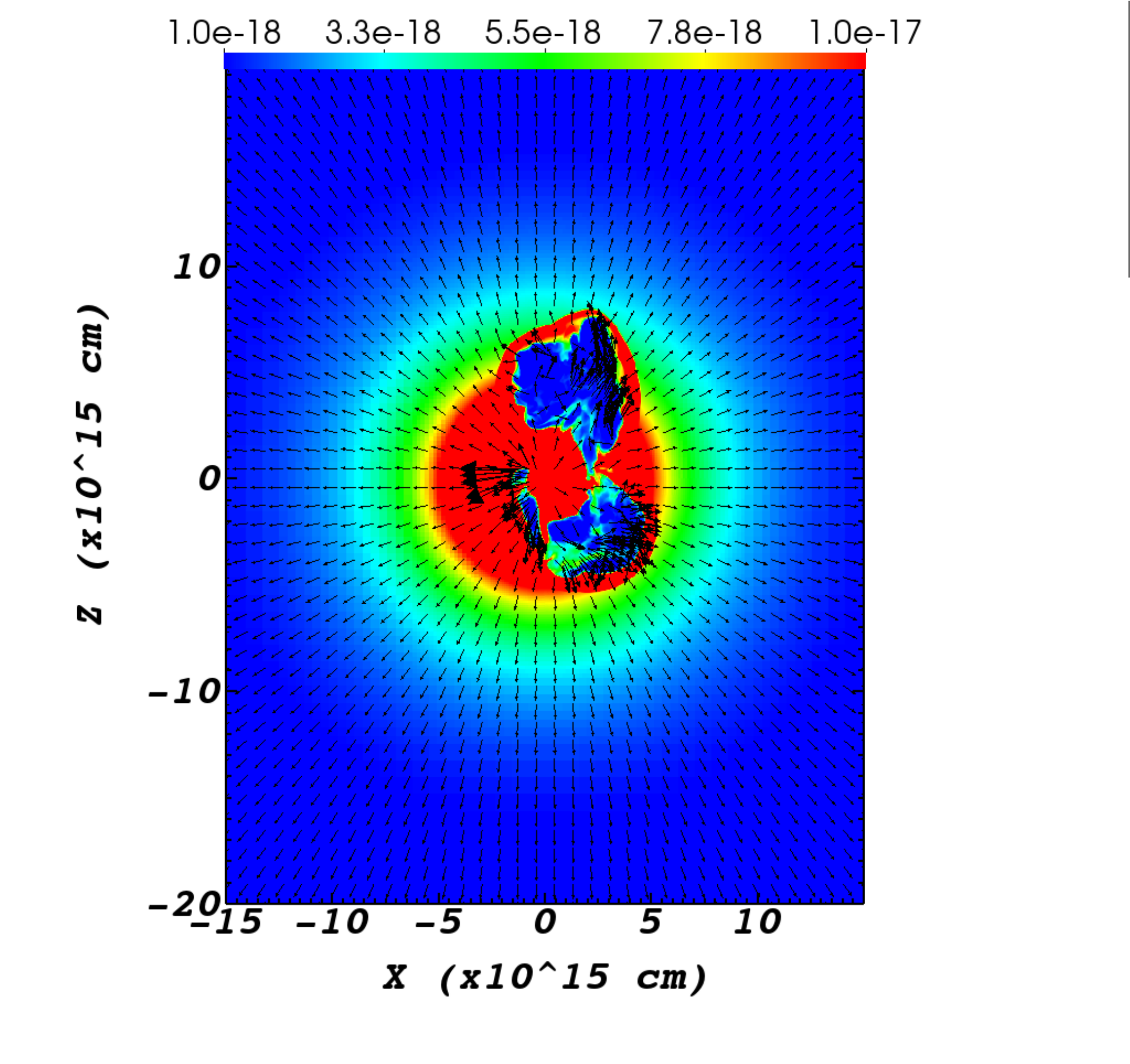}}
\subfigure[$t=72$~\yr]{\includegraphics[height=2.7in,width=2.7in,angle=0]{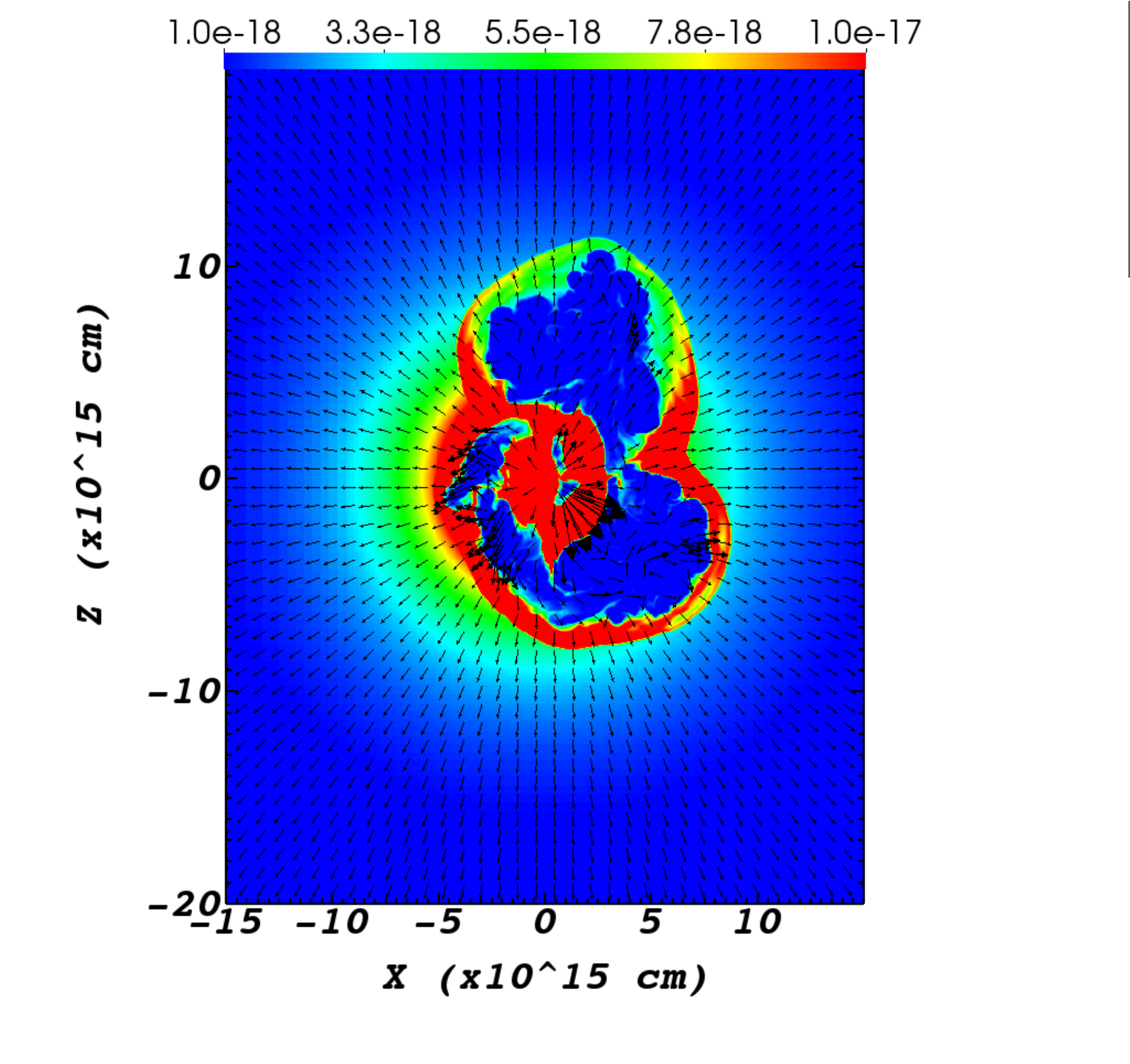}}
\subfigure[$t=109$~\yr]{\includegraphics[height=2.7in,width=2.7in,angle=0]{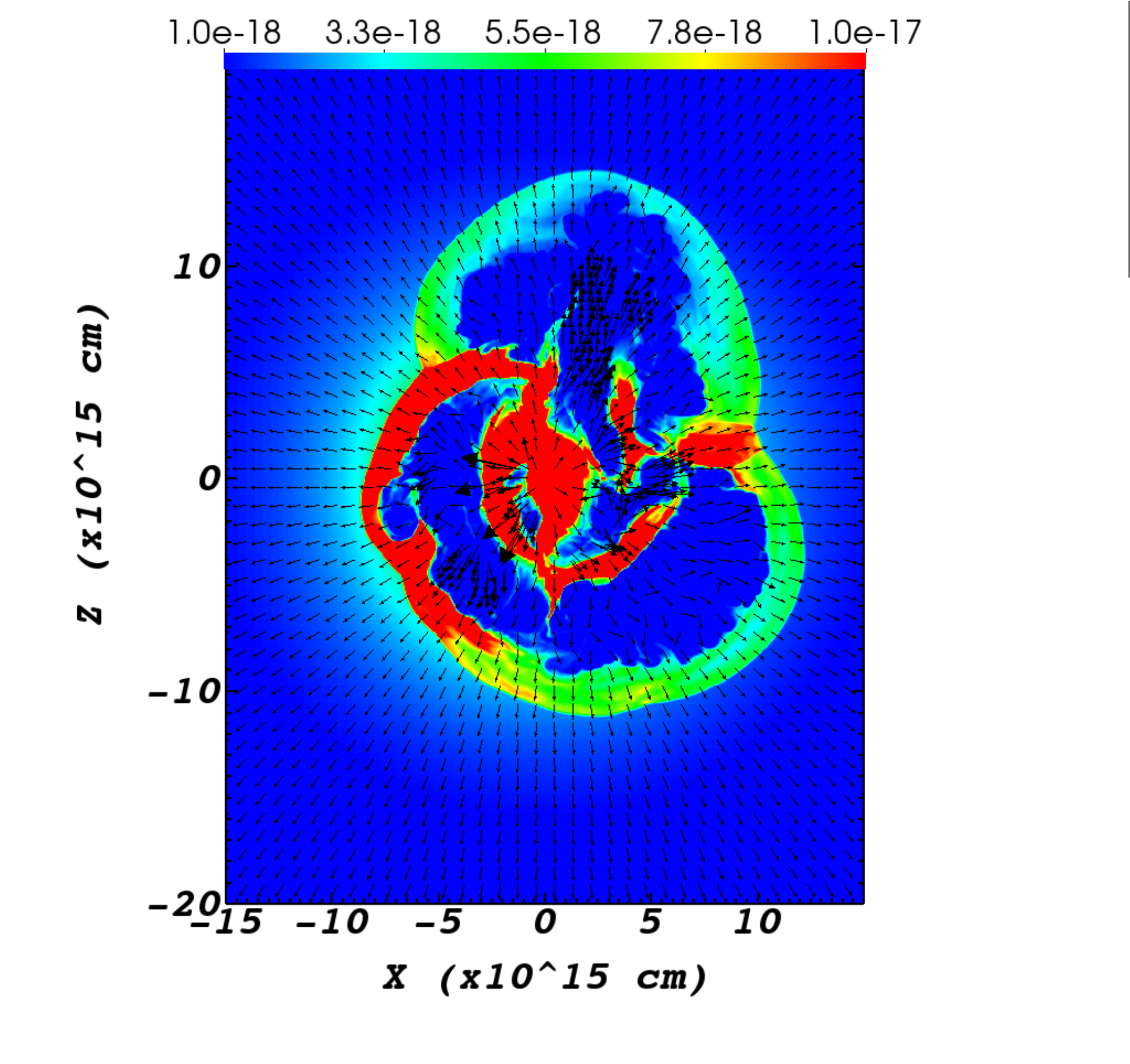}}
\subfigure[$t=130$~\yr]{\includegraphics[height=2.7in,width=2.7in,angle=0]{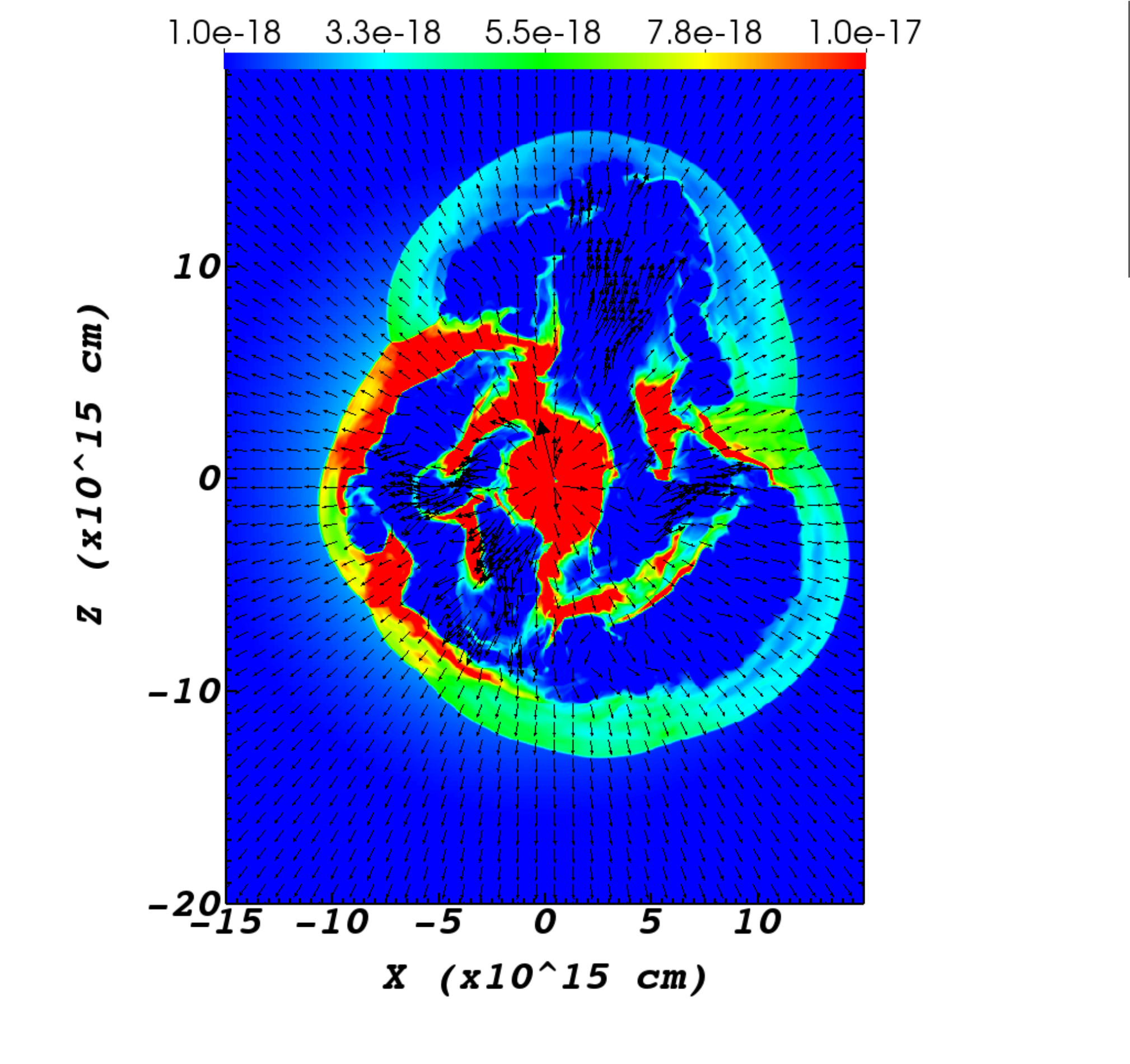}}
\caption{The density maps in the $y=0$ plane for the $\phi = 20^\circ$ case at the times $41$, $72$, $109$, and $130 \yr$.
The AGB star is at   $(x,y,z)=(0,0,-10) \AU$.
At $t=0$ the tight binary system starts to launch the jets at the coordinate $(x,y,z)=(28, 0, 0) \AU$.
The density scaling is shown in the colored bars in units of $\g \cm^{-3}$. }
 \label{fig:rho20degree}
\end{center}
\end{figure}

In Fig. \ref{fig:rho20degree} we present density maps in the $y=0$ plane for
the $\phi=20^\circ$ case (see Fig.  \ref{fig:schematic}), at four times.
In all panels the AGB star is located just below the center of the grid
at $(x,y,z)=(0,0,-10)\AU$. The source of the jets orbits the AGB star, and
its location is $(x,y,z)=(-21.5, -19.4, -18.1 )\AU$ at $t=41 \yr$, $(x,y,z)=(25.1, 13.6, -1.1 )\AU$
at $t=72 \yr$, $(x,y,z)=(-19.6, -21.5, -17.4 )\AU$ at $t=109 \yr$, and $(x,y,z)=(26.2, -11, -0.7 )\AU$ at $t=130 \yr$.
  
The main outcome of the simulation is a nebula that lacks any symmetry.
We simulate only the early interaction time, and cannot deduce the exact final
nebular shape that is observed at later time. Nonetheless, our results
clearly demonstrates the early formation of a `messy PN'.
The arrows depict the flow direction at each point, but not its value. We can notice a complicated flow structure in the low density regions that are enclosed by the dense shell. These low density regions are filled with the shocked jets' material. These regions contain little mass, and will not be observed in the visible when the object turns into a PN. The complicated flow structure in these regions, including many vortexes, forms many filaments in the dense gas; these will be observed in the PN phase. We will now turn to describe only the $\phi=30^\circ$ case.
 
\begin{figure}
\begin{center}
\subfigure[$t=41$~\yr]{\includegraphics[height=2.7in,width=2.7in,angle=0]{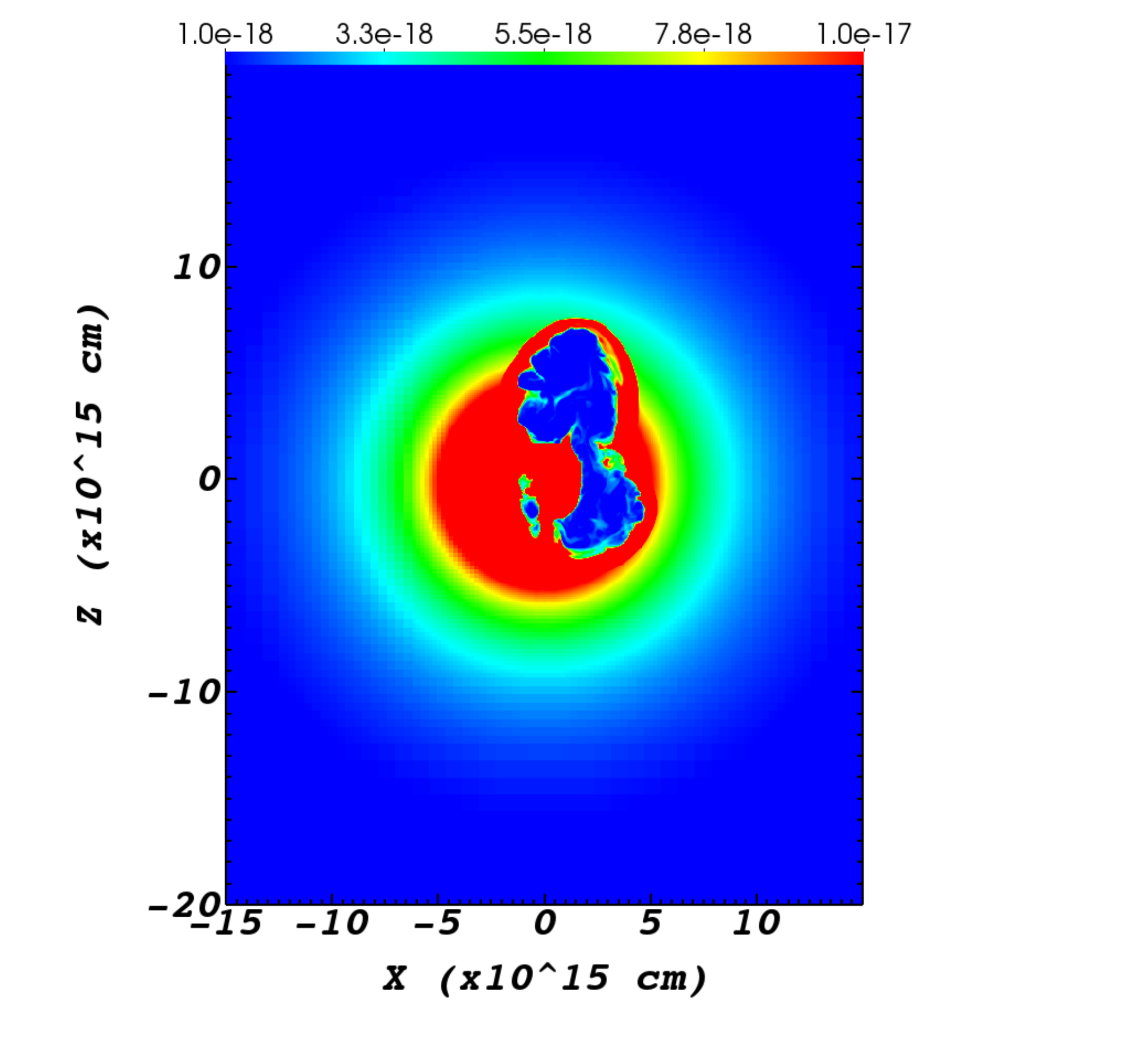}}
\subfigure[$t=72$~\yr]{\includegraphics[height=2.7in,width=2.7in,angle=0]{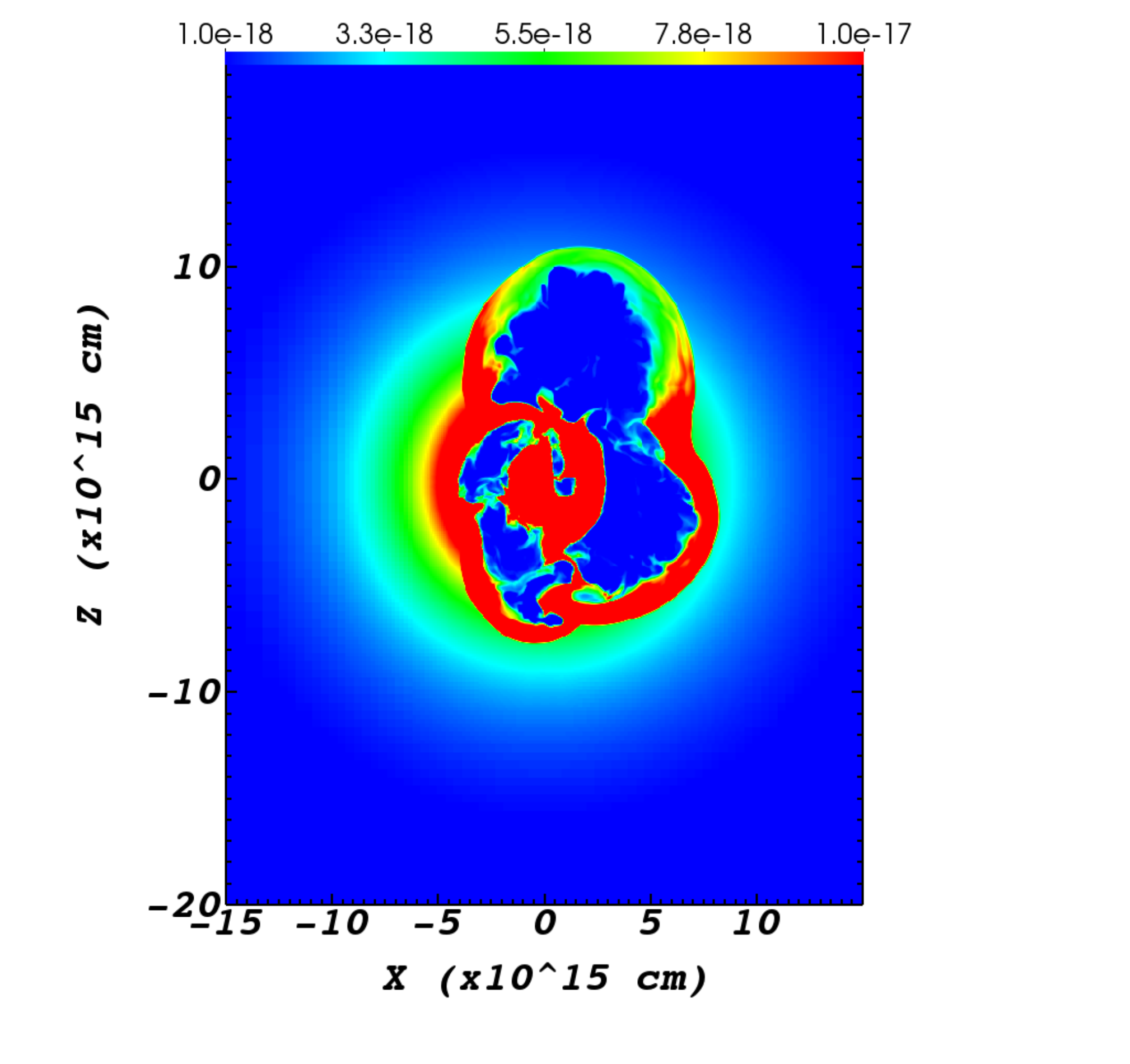}}

\subfigure[$t=109$~\yr]{\includegraphics[height=2.7in,width=2.7in,angle=0]{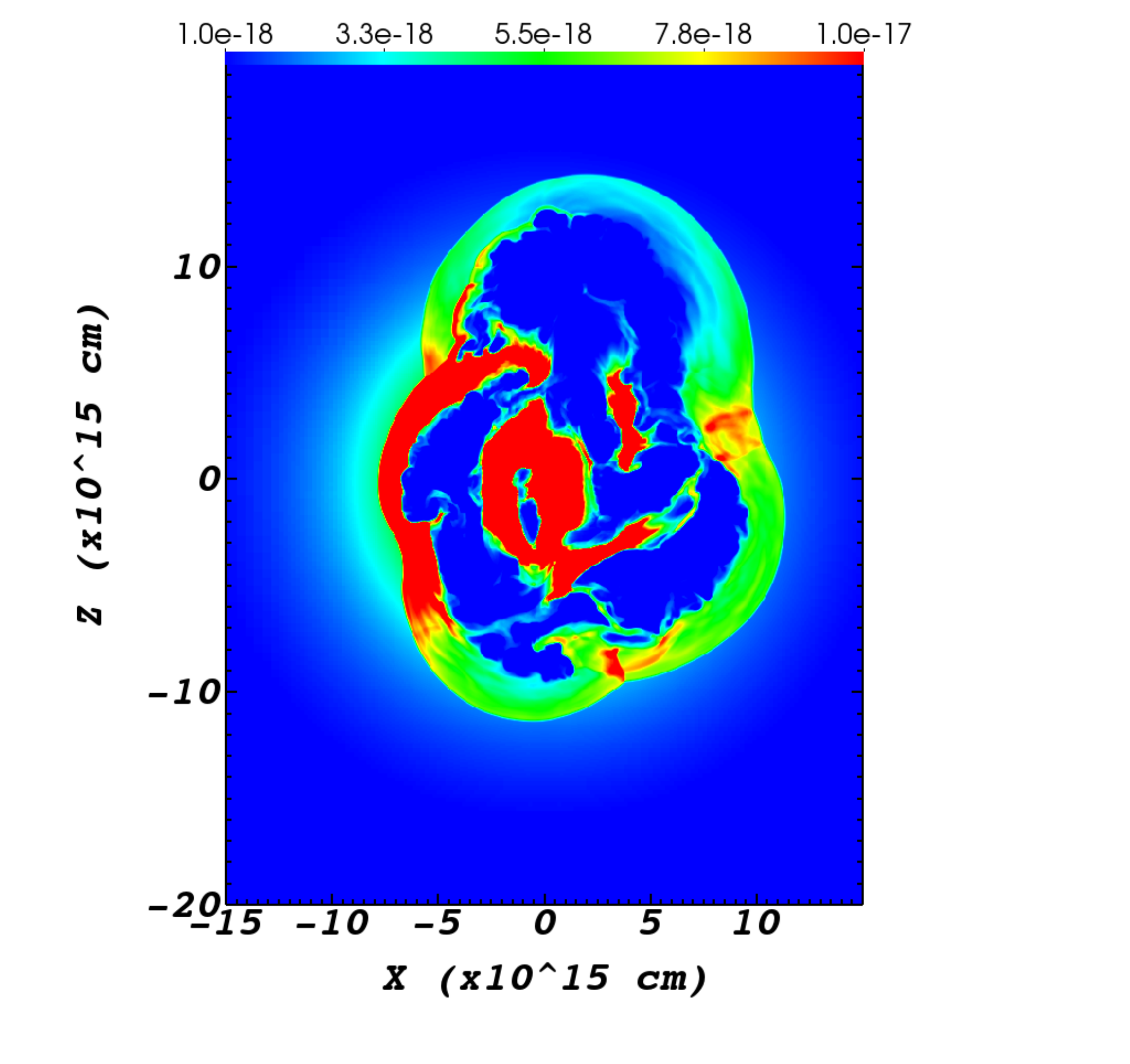}}
\subfigure[$t=130$~\yr]{\includegraphics[height=2.7in,width=2.7in,angle=0]{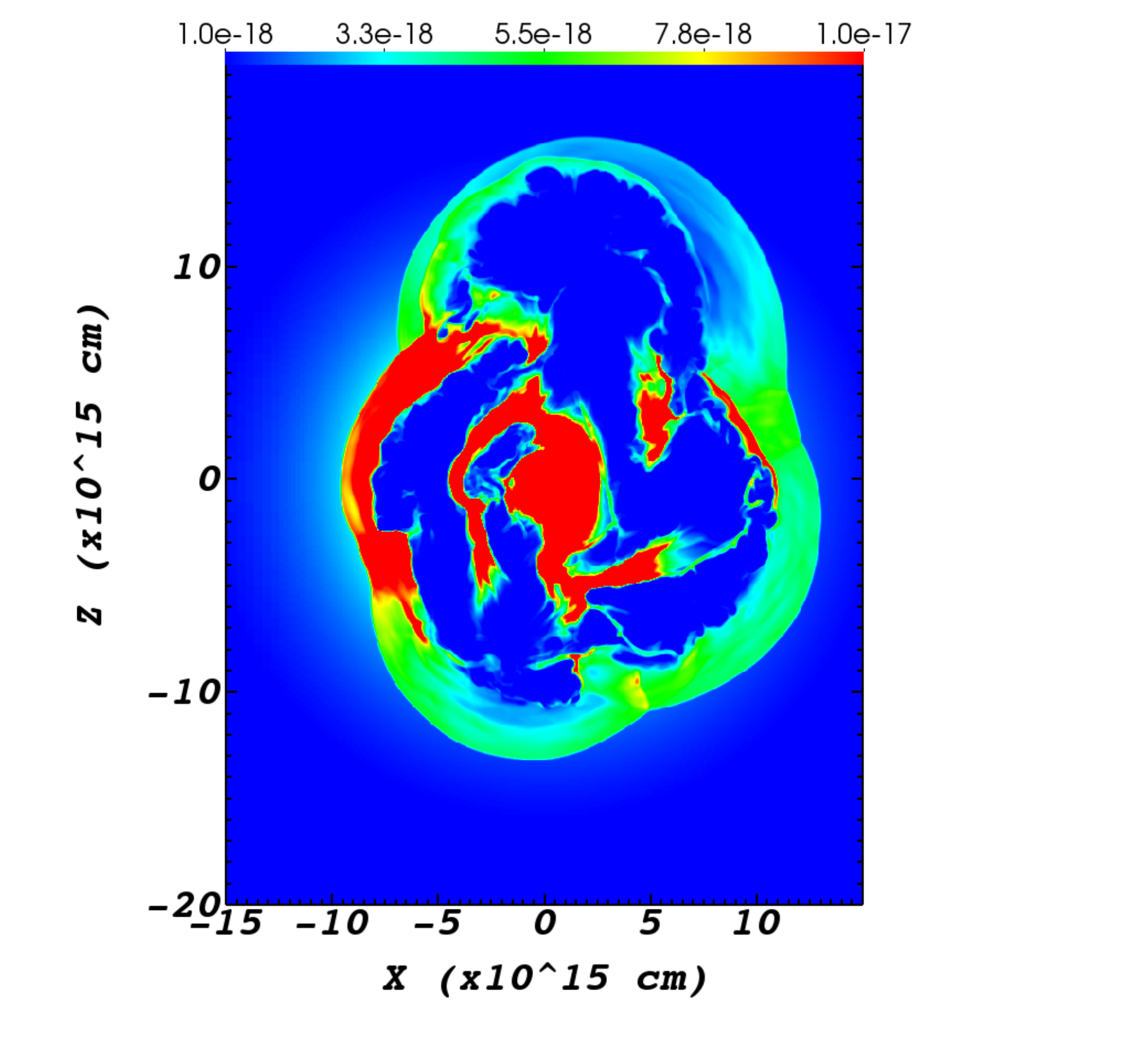}}

\caption{Like figure \ref{fig:rho20degree}, but for the case $\phi = 30^\circ$. At $t=0$ the AGB star is at $(x,y,z)=(0,0,-15) \AU$ and the tight binary system starts to launch the jets at the coordinate $(x,y,z)=(26, 0, 0) \AU$.}
 \label{fig:rho30degree}
\end{center}
\end{figure}

In Fig. \ref{fig:rho30degree} we present the density maps for the $\phi=30^\circ$ simulation.
Now the AGB star is located below the center of the grid at $(x,y,z)=(0,0,-15)\AU$, and the  source of the jets that orbits the AGB star is located at $(x,y,z)=(-19.8, -19.4, -26.4)\AU$ at $t=41 \yr$, $(x,y,z)=(23.2, 13.6, -1.6)\AU$ at $t=72 \yr$, $(x,y,z)=(-18.2, -21.5, -25.5)\AU$ at $t=109 \yr$, and $(x,y,z)=(24.2, -11, -1)\AU$ at $t=130 \yr$.
The results are not much different from those for the $\phi=20^\circ$ case presented in Fig. \ref{fig:rho20degree}.
The two main morphological features are a 'messy' shape and the formation of a filamentary structure.

\begin{figure}
\begin{center}

\subfigure[$t=41$~\yr]{\includegraphics[height=2.7in,width=2.7in,angle=0]{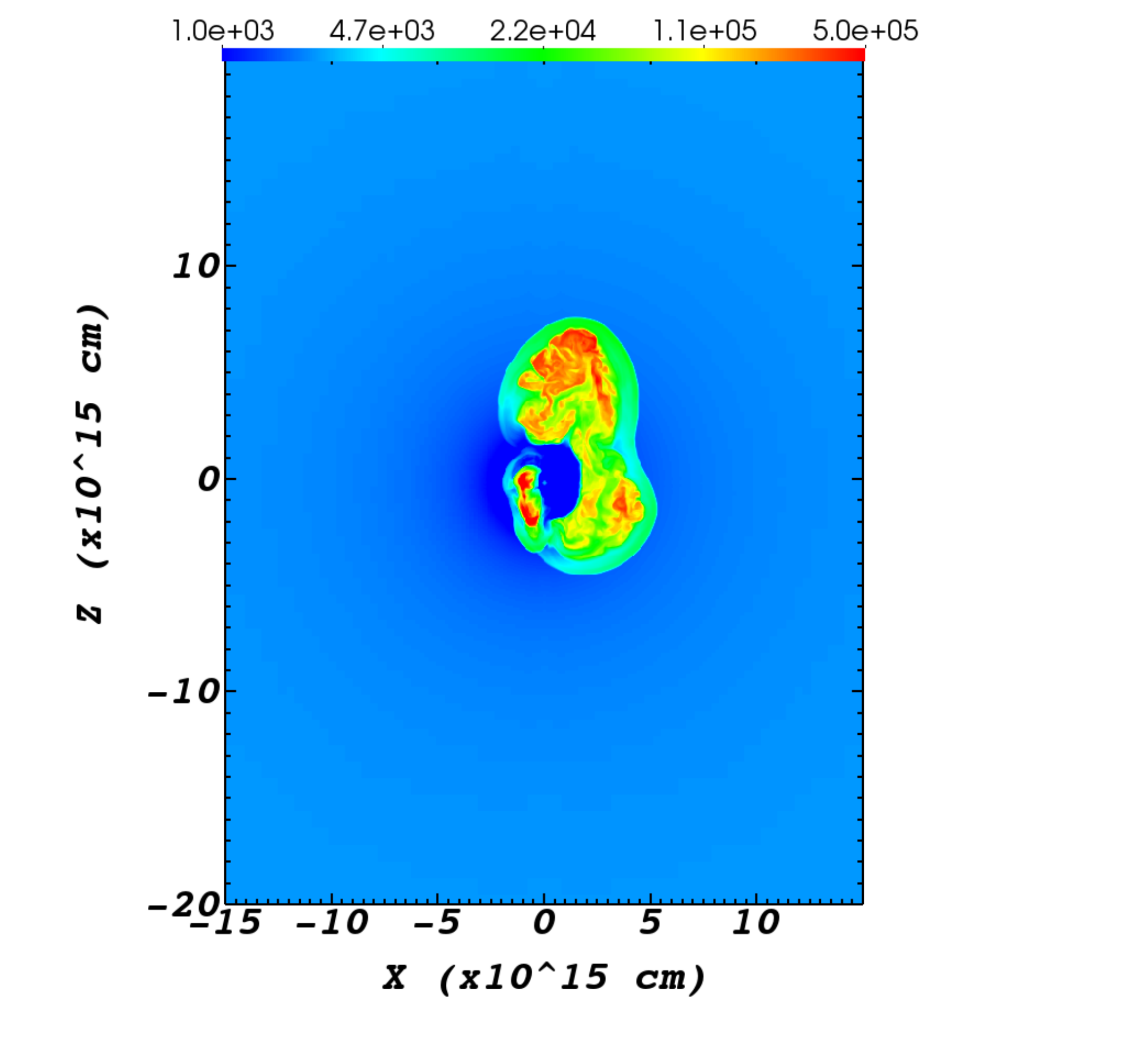}}
\subfigure[$t=72$~\yr]{\includegraphics[height=2.7in,width=2.7in,angle=0]{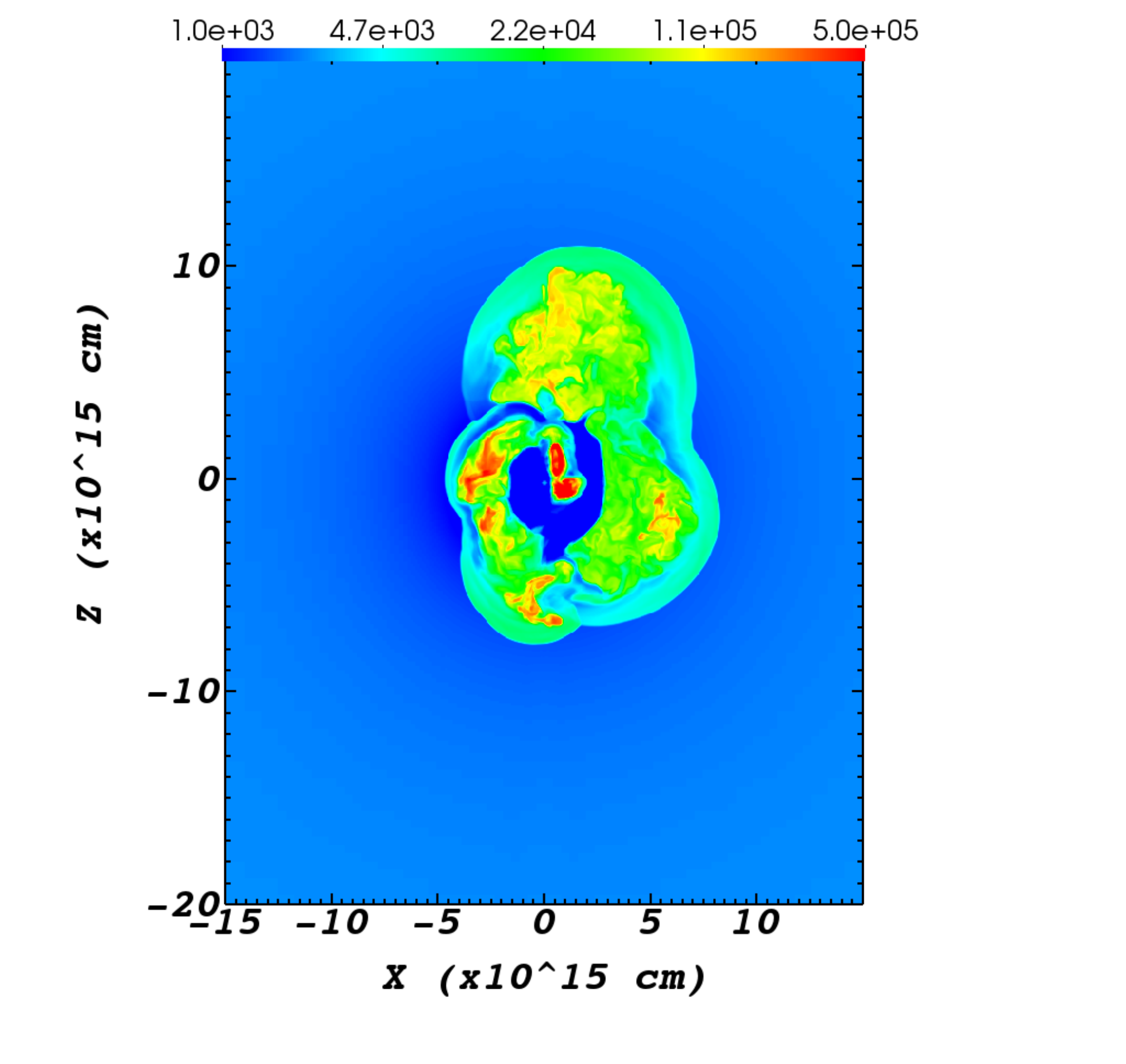}}

\subfigure[$t=109$~\yr]{\includegraphics[height=2.7in,width=2.7in,angle=0]{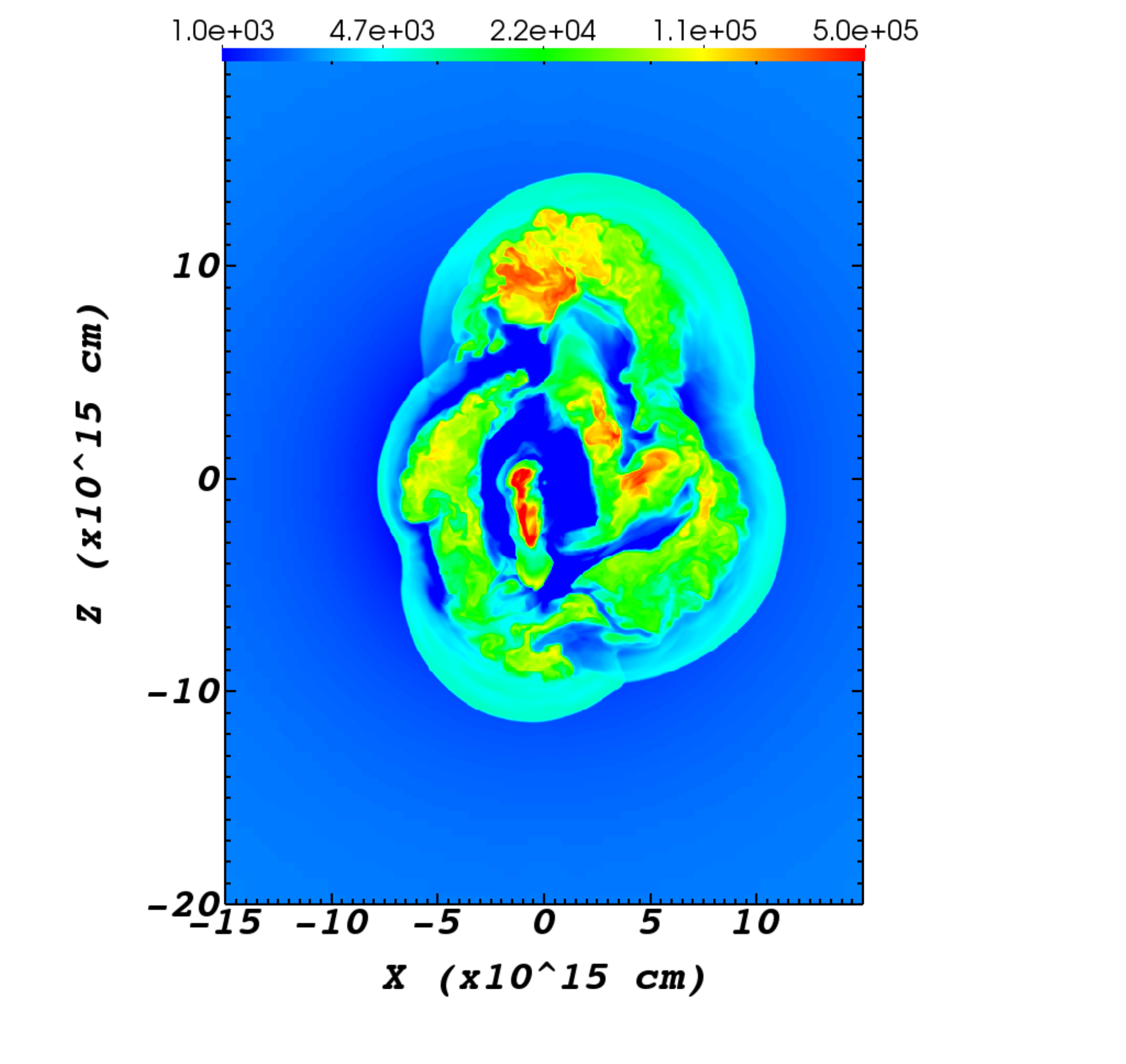}}
\subfigure[$t=130$~\yr]{\includegraphics[height=2.7in,width=2.7in,angle=0]{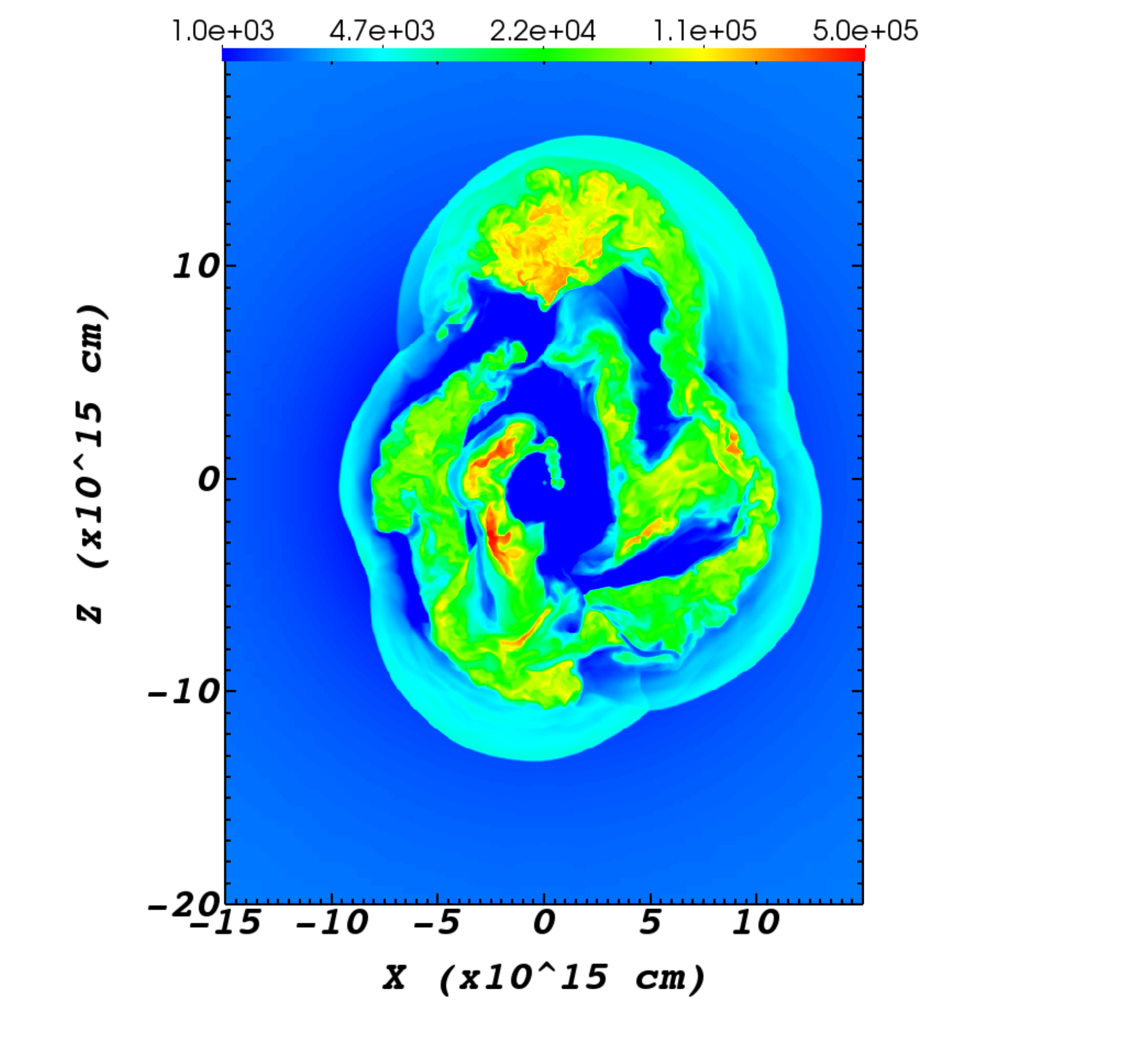}}
\caption{Like figure \ref{fig:rho30degree}, but for temperatures maps (in Kelvin).}
 \label{fig:temp30degree}
\end{center}
\end{figure}

We present the temperature maps (in Kelvin) in Fig. \ref{fig:temp30degree}, corresponding to the same planes and times as in Fig. \ref{fig:rho30degree}. The jets are shocked to temperatures of few millions degrees Kelvin. The post-shock gas suffers a substantial adiabatic cooling. Nonetheless, the high temperatures regions in the maps correspond to post-shock jets' material. These are regions of low density, that when surrounded by dense AGB gas are termed bubbles. 

\begin{figure}[h!]
\begin{center}

\subfigure[$\theta=30^\circ$]{\includegraphics[height=3.in,width=3.in,angle=0]{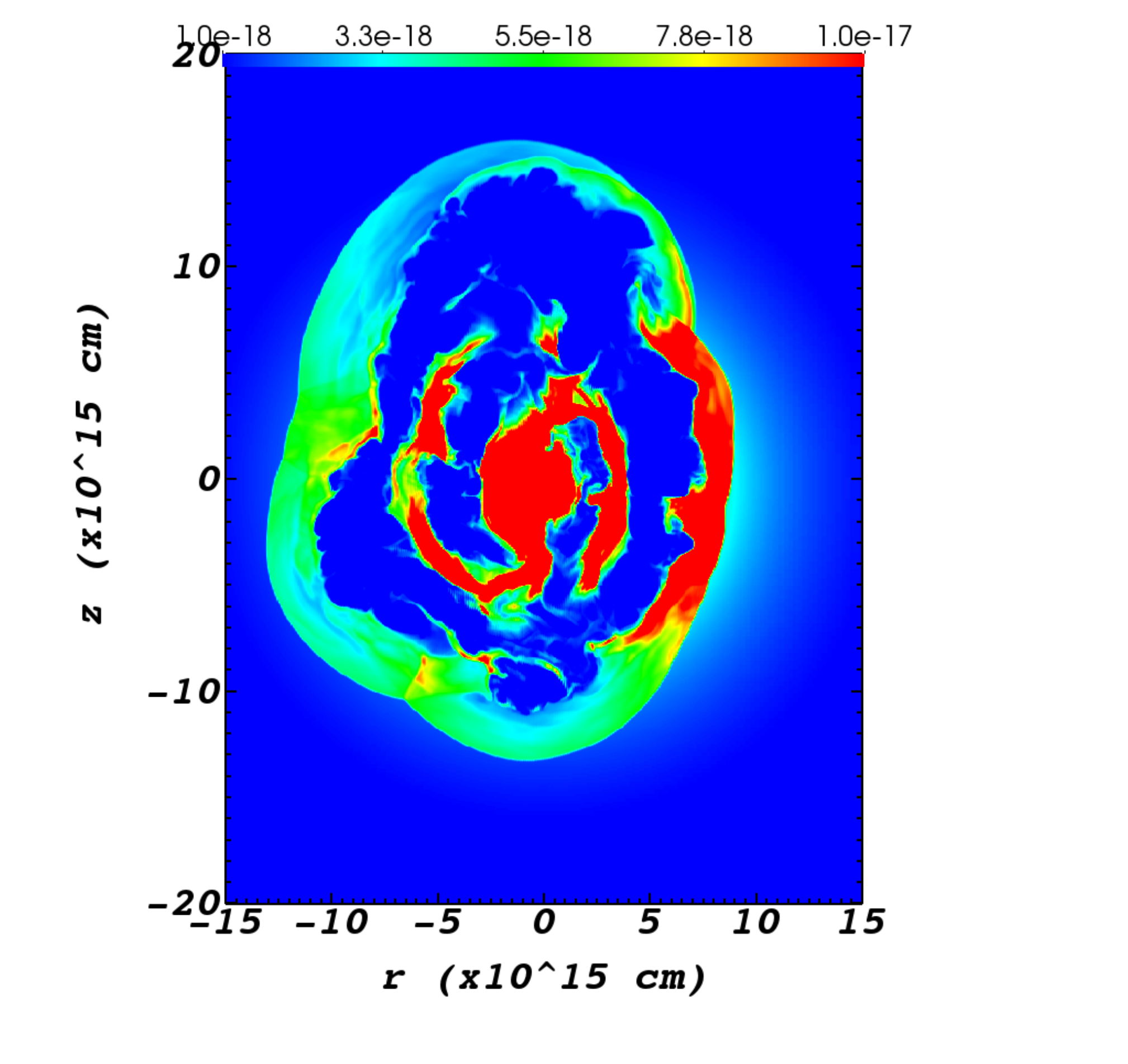}}
\subfigure[$\theta=60^\circ$]{\includegraphics[height=3.in,width=3.in,angle=0]{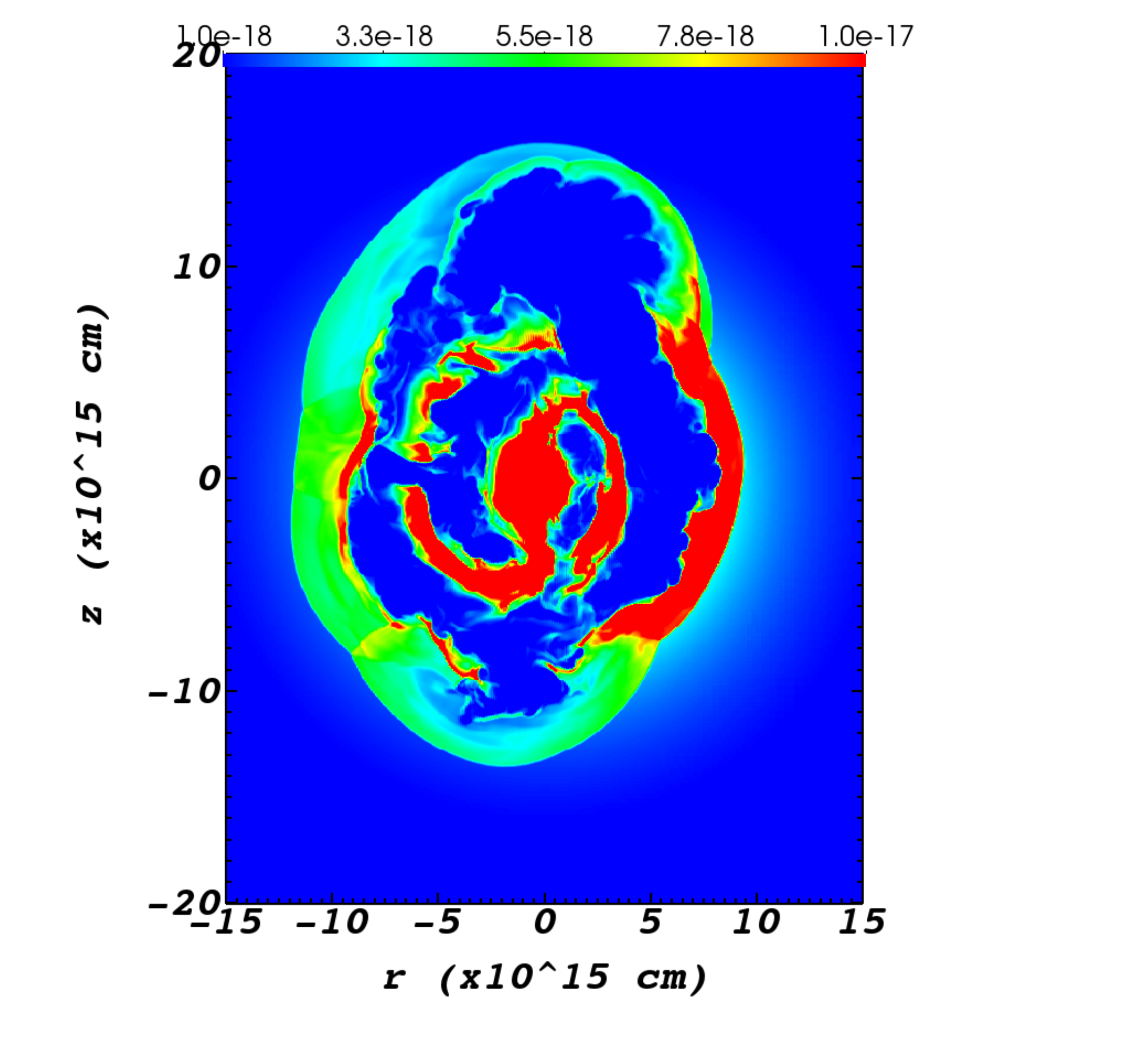}}

\subfigure[$\theta=90^\circ$]{\includegraphics[height=3.in,width=3.in,angle=0]{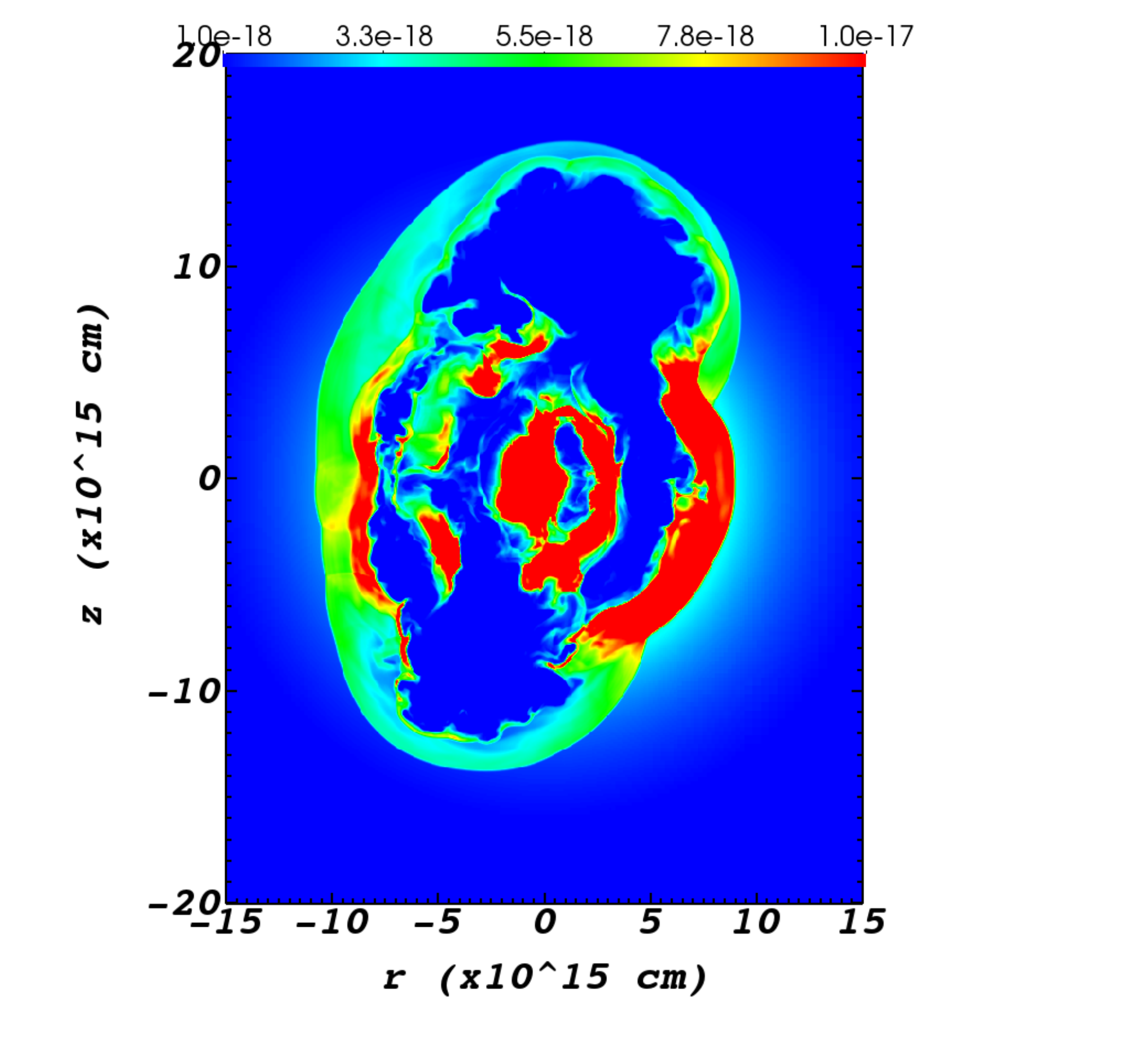}}
\subfigure[$\theta=120^\circ$]{\includegraphics[height=3.in,width=3.in,angle=0]{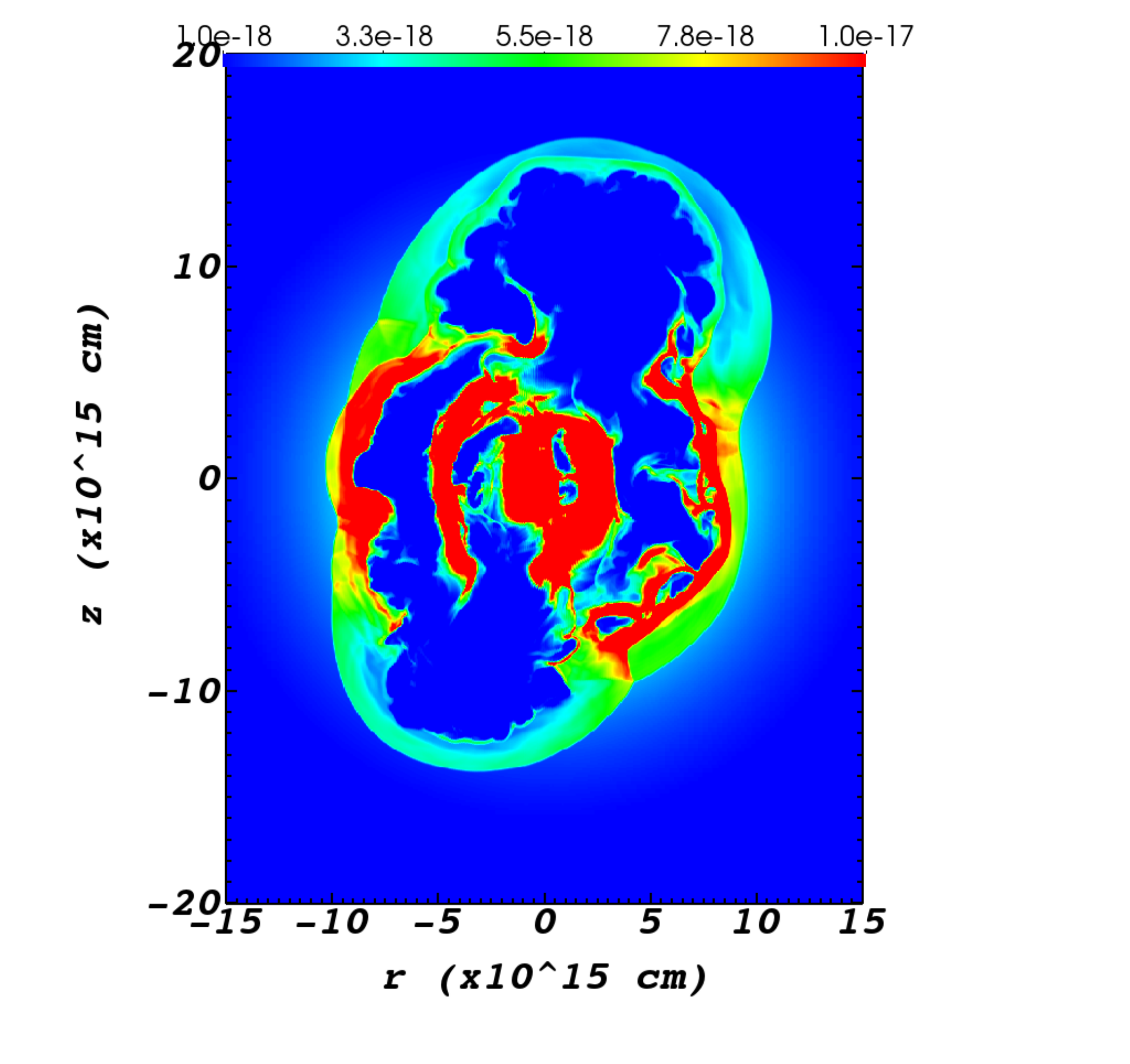}}

\caption{Density maps at $t=130~\yr$ and in four different planes perpendicular to the $x-y$ plane for the run presented in Fig.  \ref{fig:rho30degree}.
The angle $\theta$ is the angle between the shown plane and the $x-z$ plane shown in Fig. \ref{fig:rho30degree}.
In these panels the horizontal axis is not the $x$ axis of the grid as used in the previous figures, but it is rather $r=(x^2 + y^2)^{1/2}$.}
 \label{fig:cutsxz}
\end{center}
\end{figure}

\begin{figure}[h!]
\begin{center}

\subfigure[$\theta=30^\circ$]{\includegraphics[height=3.in,width=3.in,angle=0]{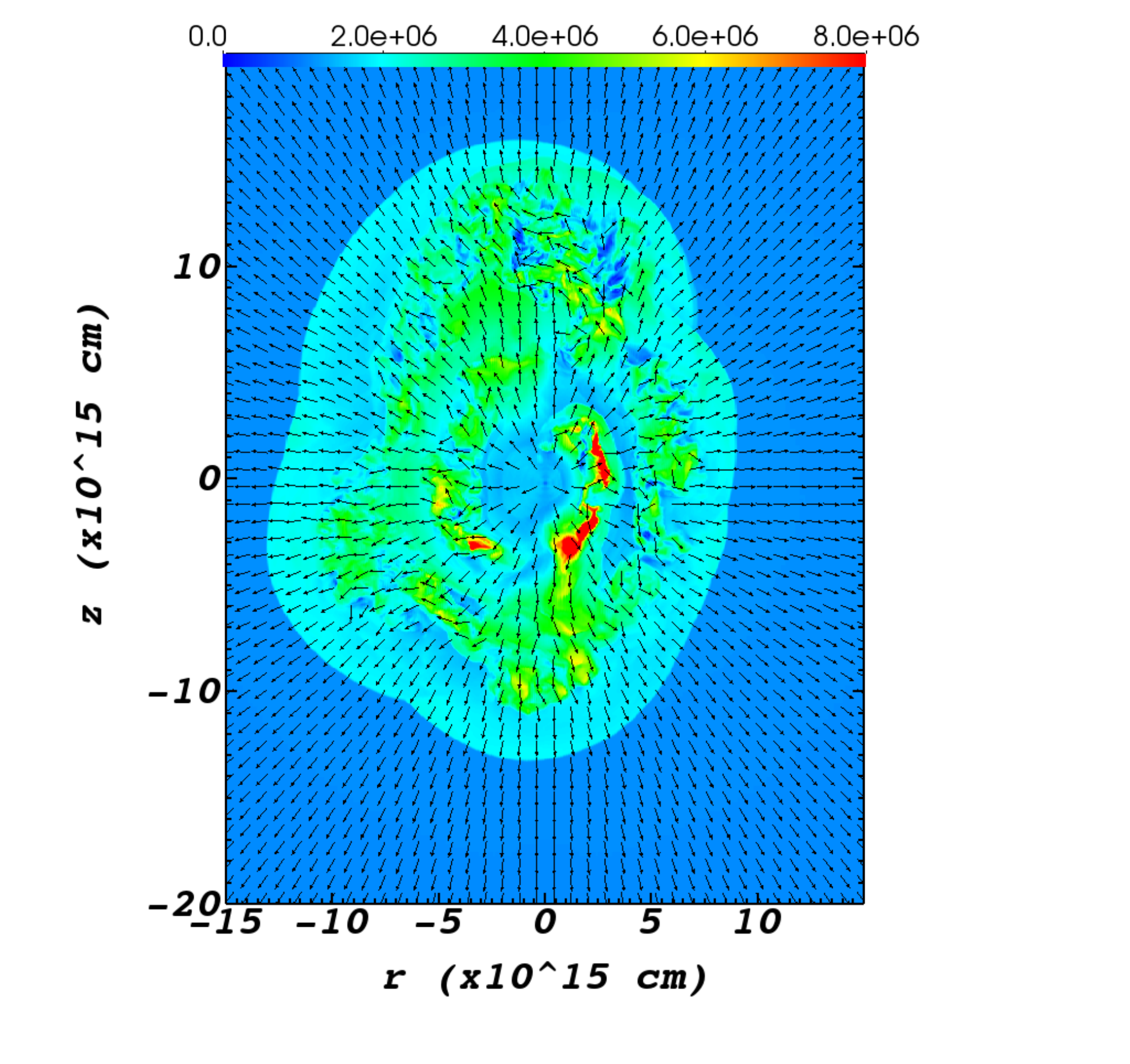}}
\subfigure[$\theta=60^\circ$]{\includegraphics[height=3.in,width=3.in,angle=0]{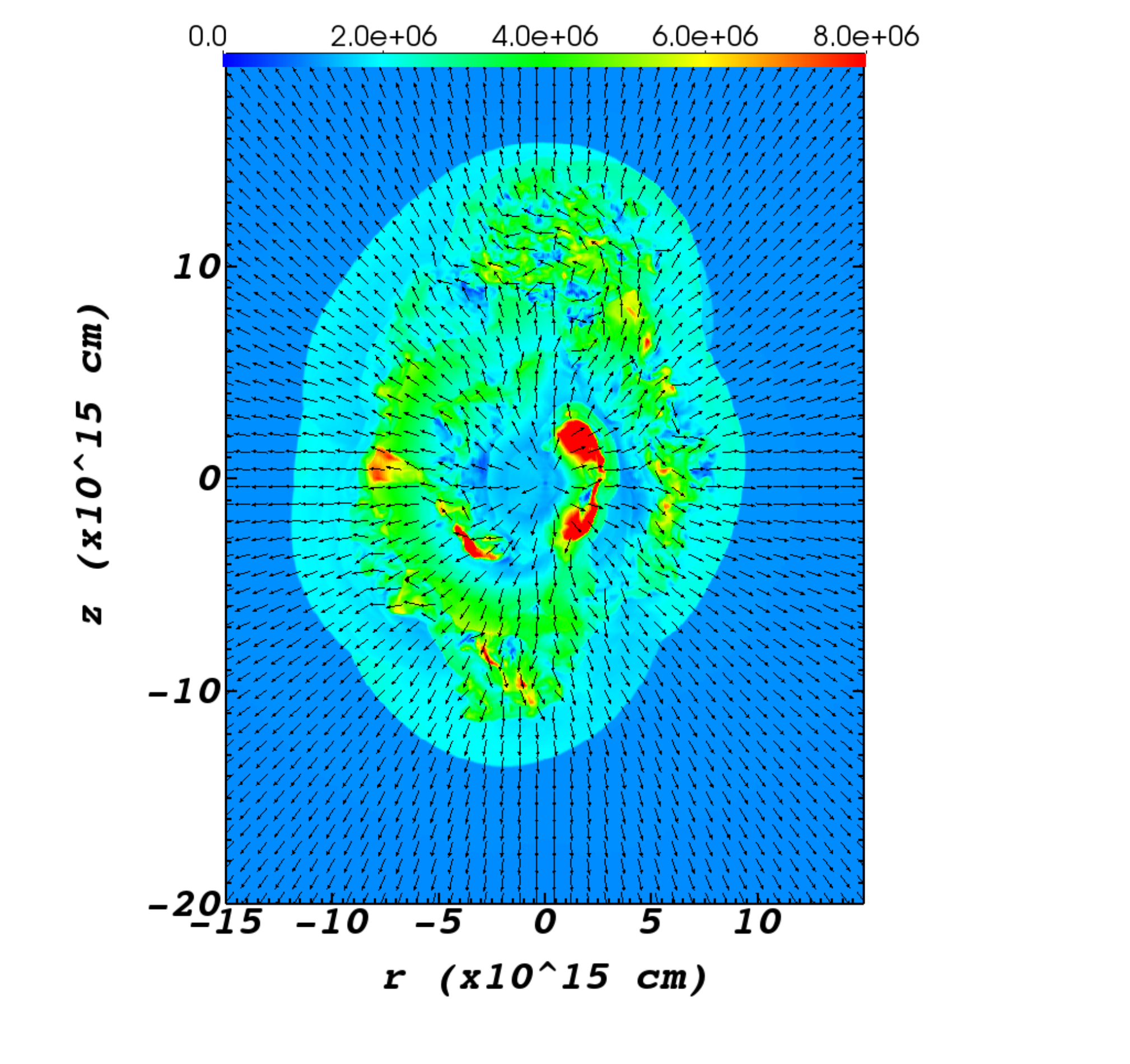}}

\subfigure[$\theta=90^\circ$]{\includegraphics[height=3.in,width=3.in,angle=0]{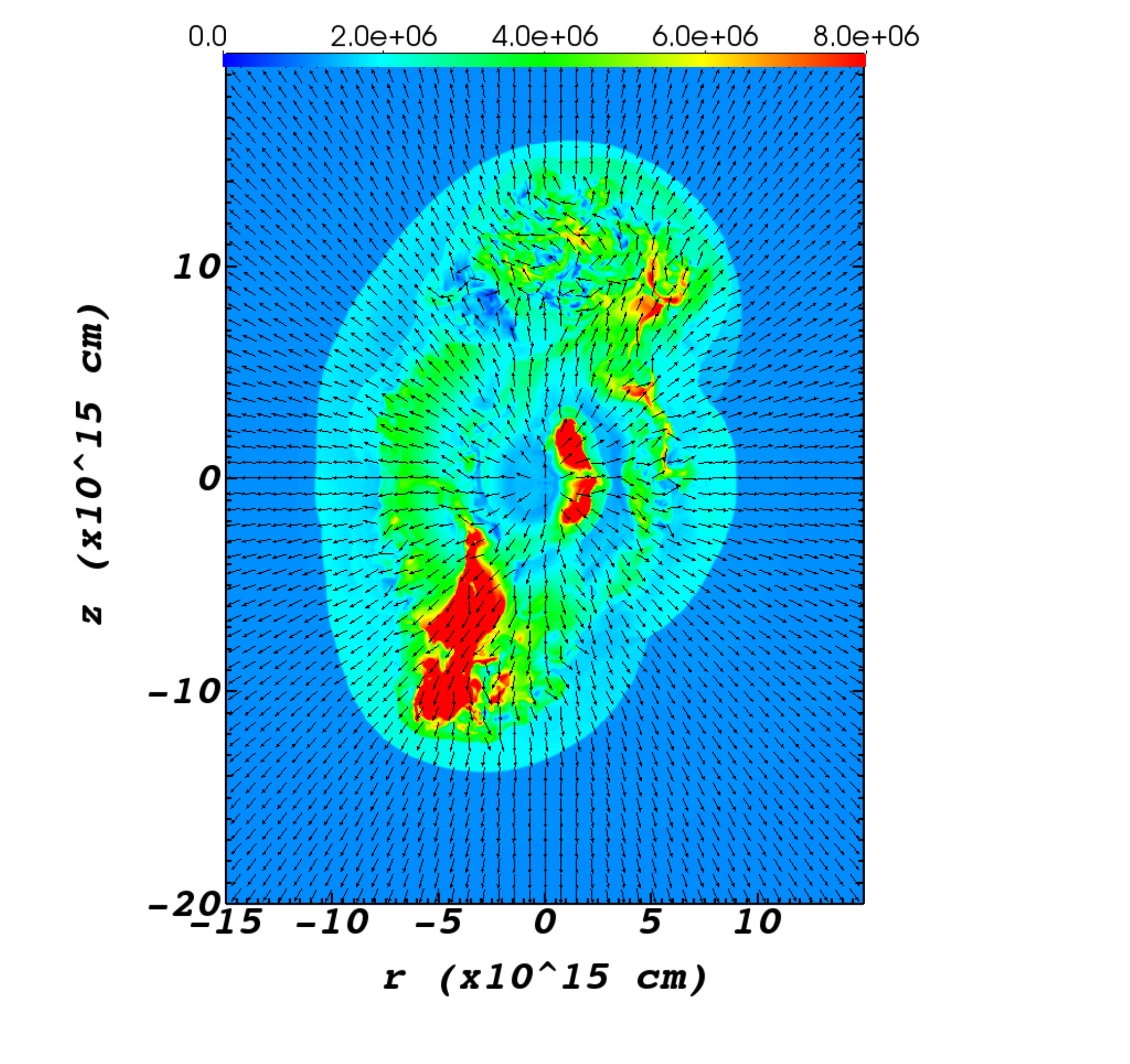}}
\subfigure[$\theta=120^\circ$]{\includegraphics[height=3.in,width=3.in,angle=0]{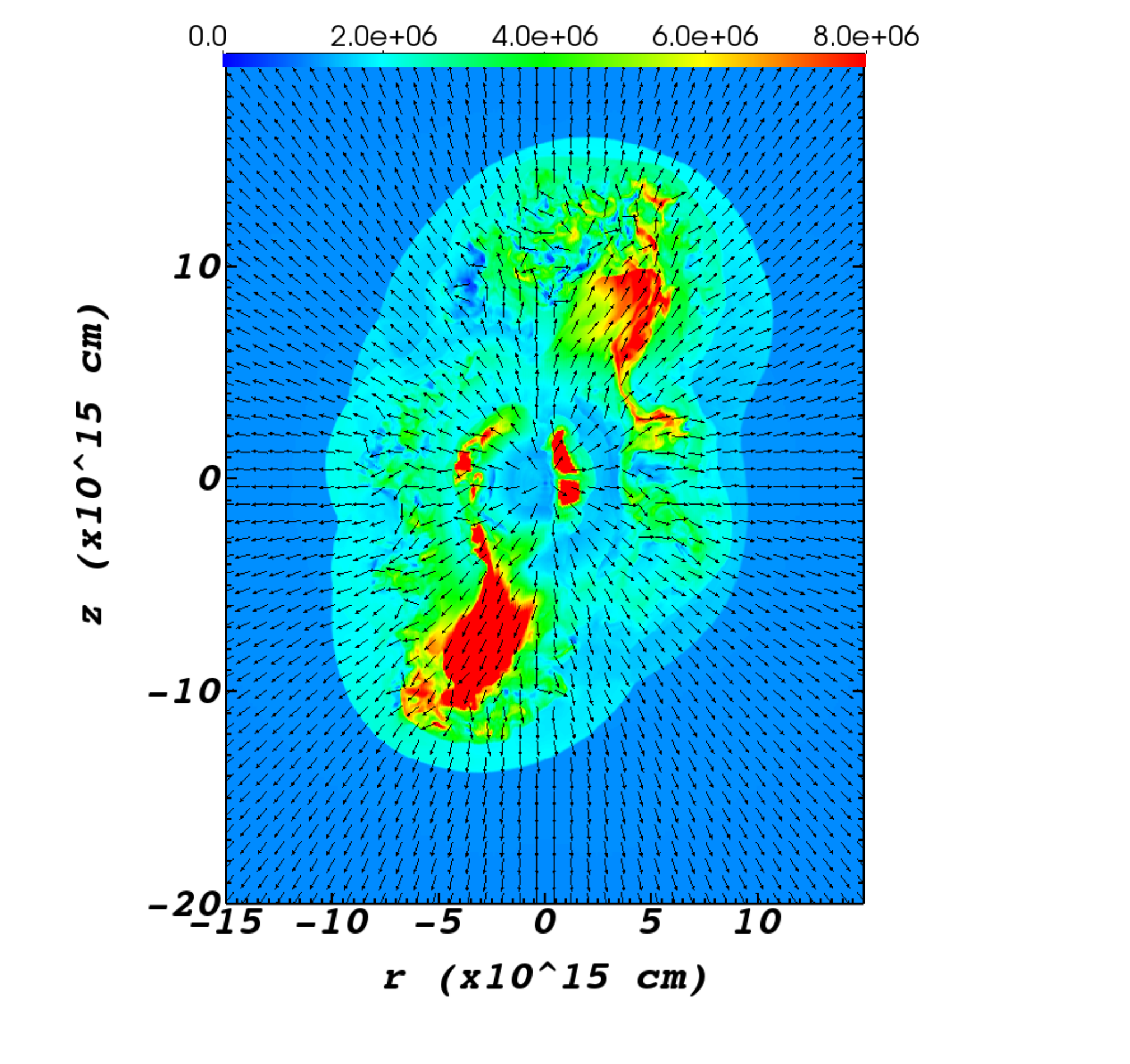}}

\caption{Velocity magnitude (color coding in $\cm \s^{-1}$) and direction (given by arrows), in the same 4 planes as
in Fig. \ref{fig:cutsxz}, and for the same run and time.}
 \label{fig:cuts_vel}
\end{center} 
\end{figure}

In Fig. \ref{fig:cutsxz} we present the density maps (again, for the $\phi=30^\circ$ case) at $t=130~\yr$ and in four different planes perpendicular to the $x-y$ plane. The angle $\theta$ is the angle between the plane and the $x-z$ plane shown in Fig. \ref{fig:rho30degree}. In these panels the horizontal axis is not the $x$ axis of the grid that is used in the previous figures, but it is actually $r=(x^2 + y^2)^{1/2}$.
In Fig. \ref{fig:cuts_vel} we plot the velocity magnitude, $v = (v_x^2+v_y^2+v_z^2)^{1/2}$, given in units of $\cm \s^{-1}$ by the color bar, and show the velocity direction with arrows, in the same four planes.
The four panels in these two figures emphasize the lack of any symmetry in the outflow, i.e., a `messy nebula', as each panel presents a different density structure. The panels show also the presence of many dense filaments in the nebula. 
 
\begin{figure}[h!]
\begin{center}
\subfigure{\includegraphics[height=2.7in,width=2.7in,angle=0]{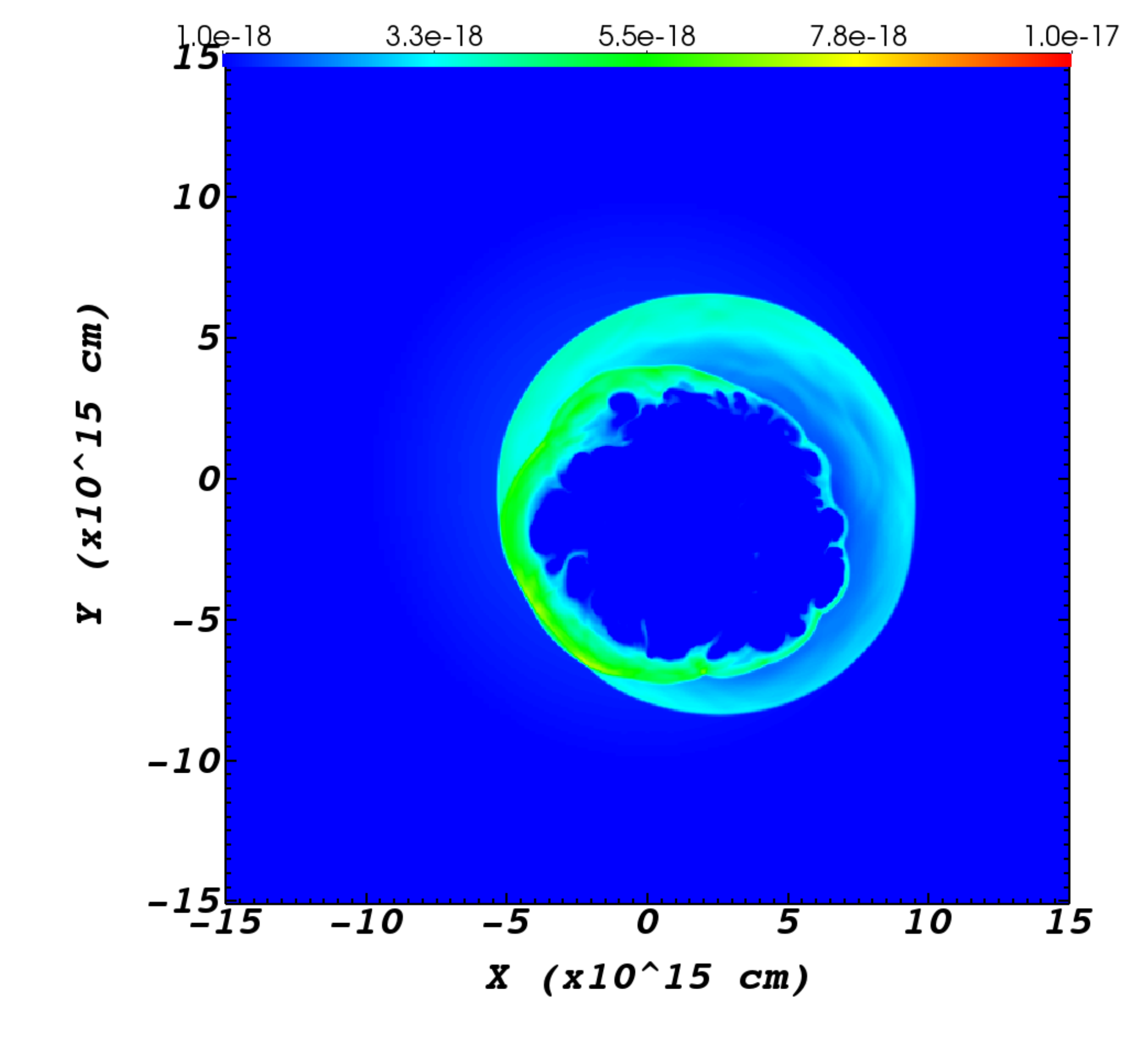}}
\subfigure{\includegraphics[height=2.7in,width=2.7in,angle=0]{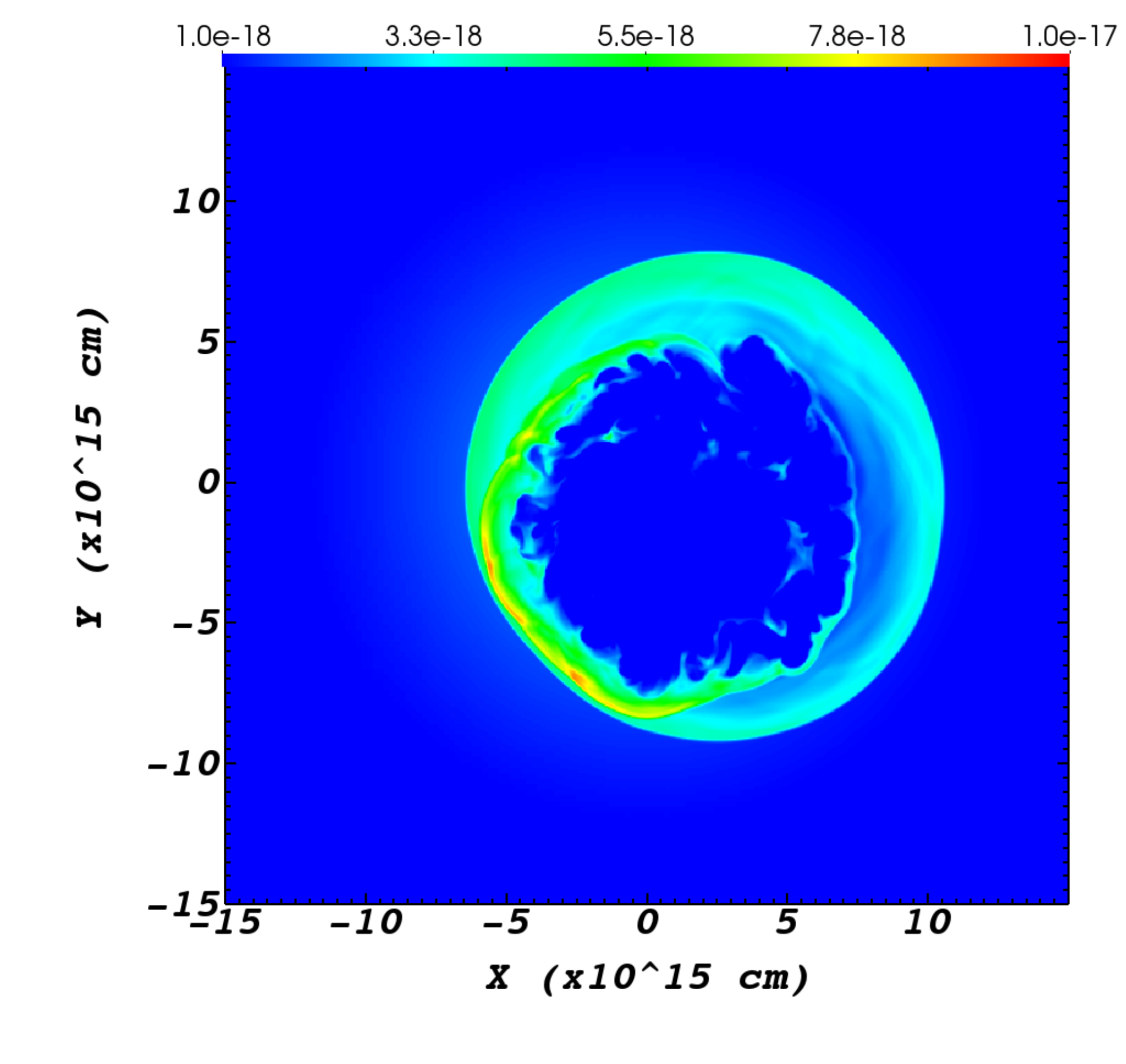}}
\vskip -1. cm
\subfigure{\includegraphics[height=2.7in,width=2.7in,angle=0]{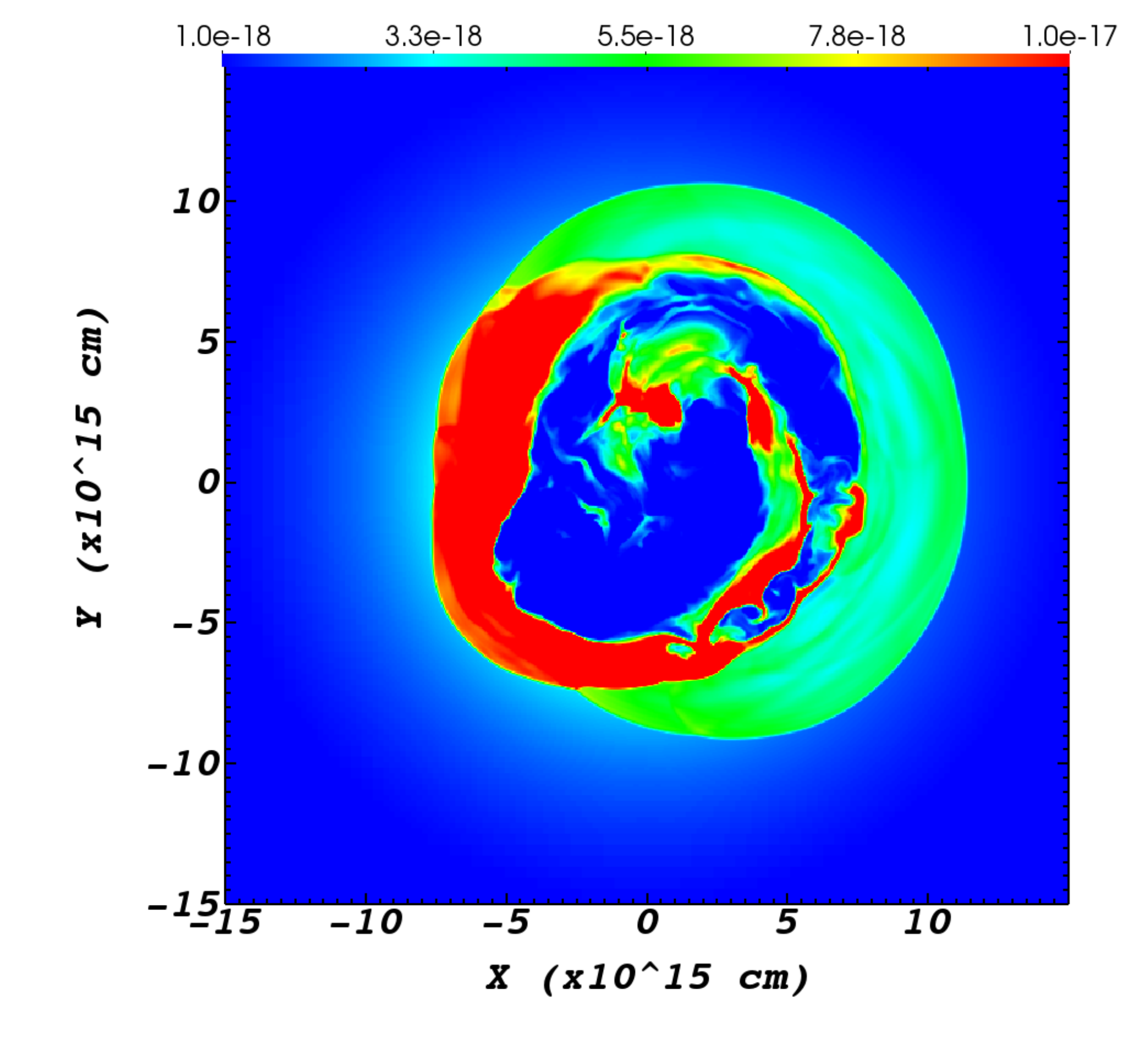}}
\subfigure{\includegraphics[height=2.7in,width=2.7in,angle=0]{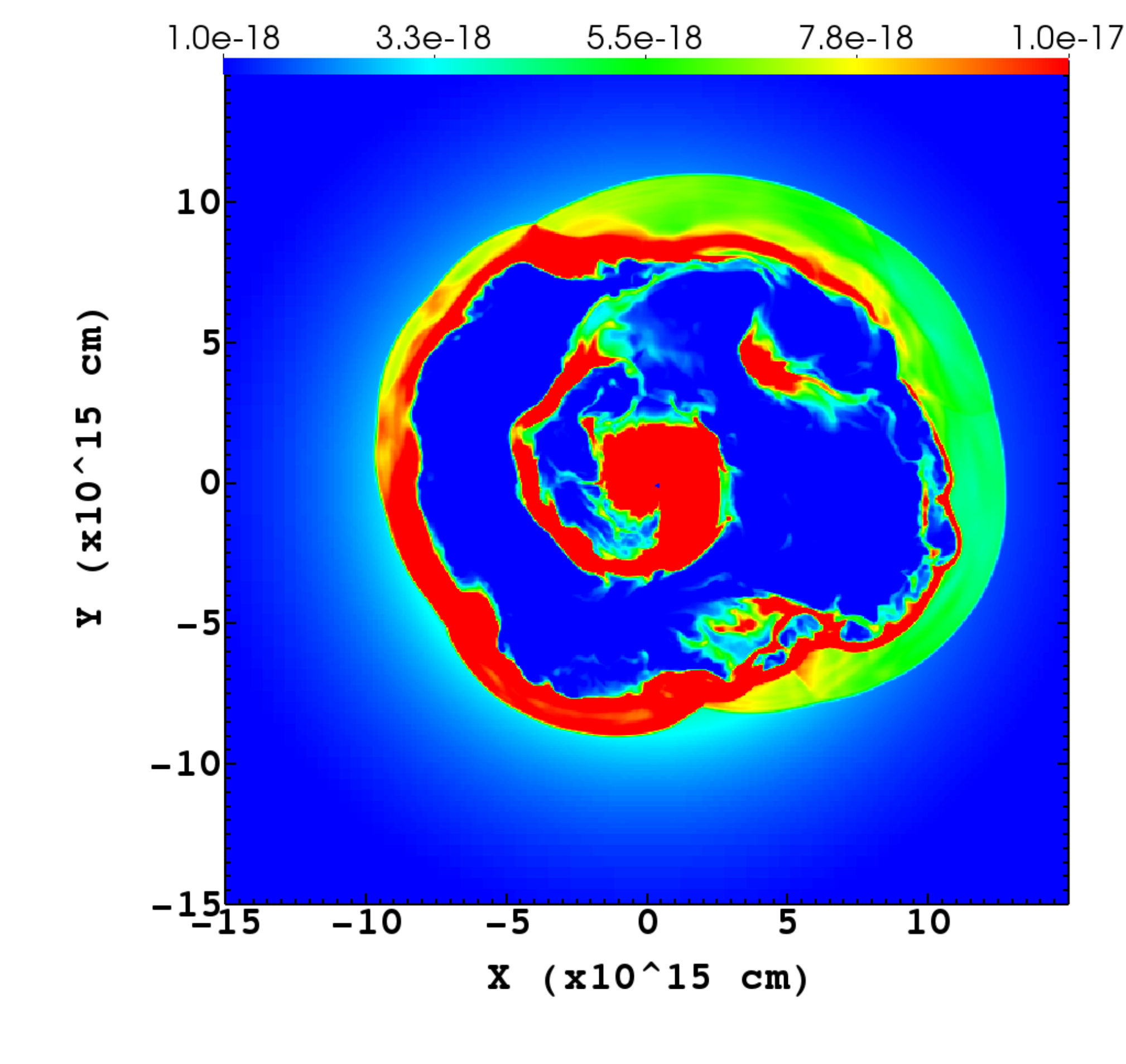}}
\vskip -1. cm
\subfigure{\includegraphics[height=2.7in,width=2.7in,angle=0]{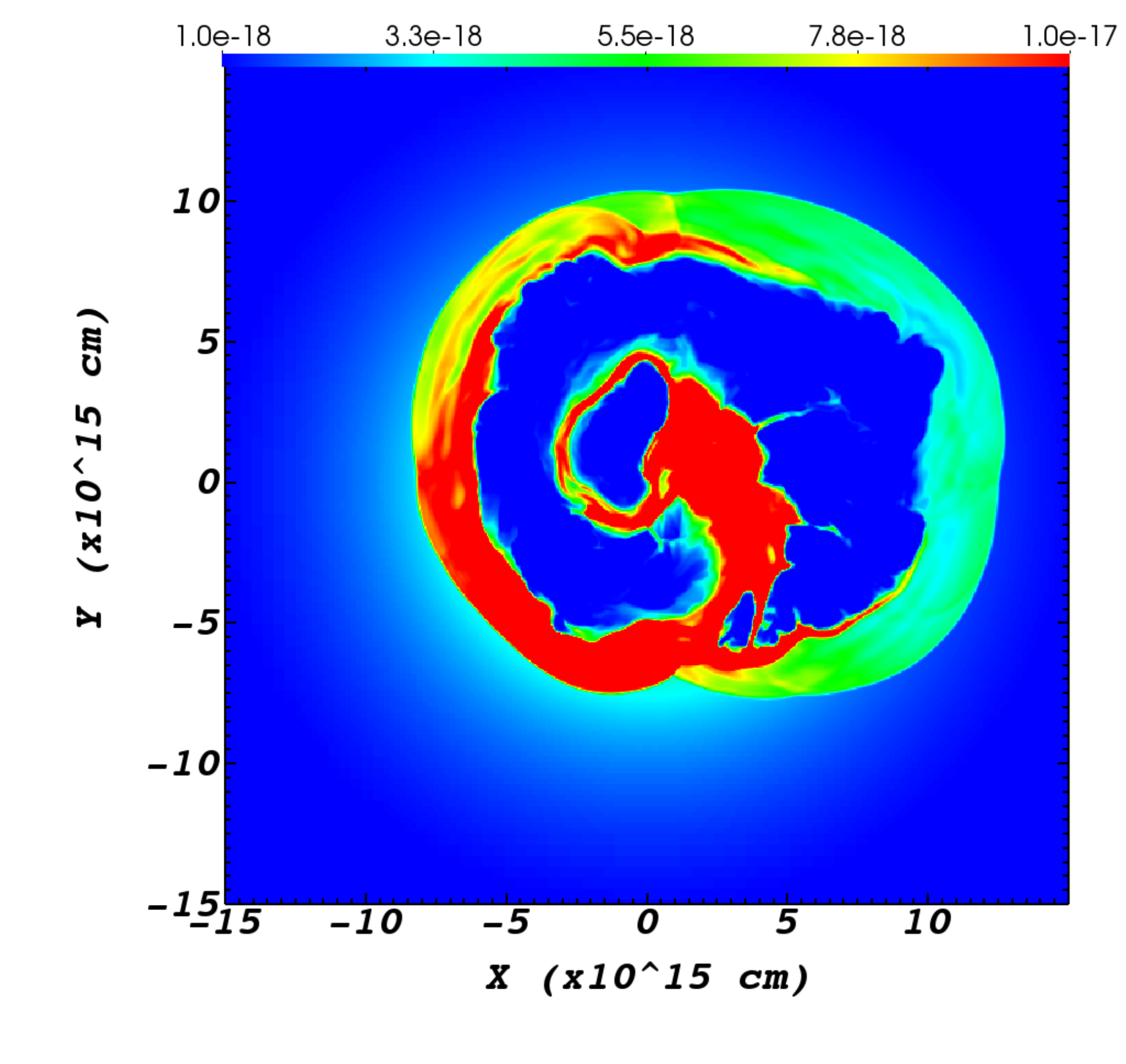}}
\subfigure{\includegraphics[height=2.7in,width=2.7in,angle=0]{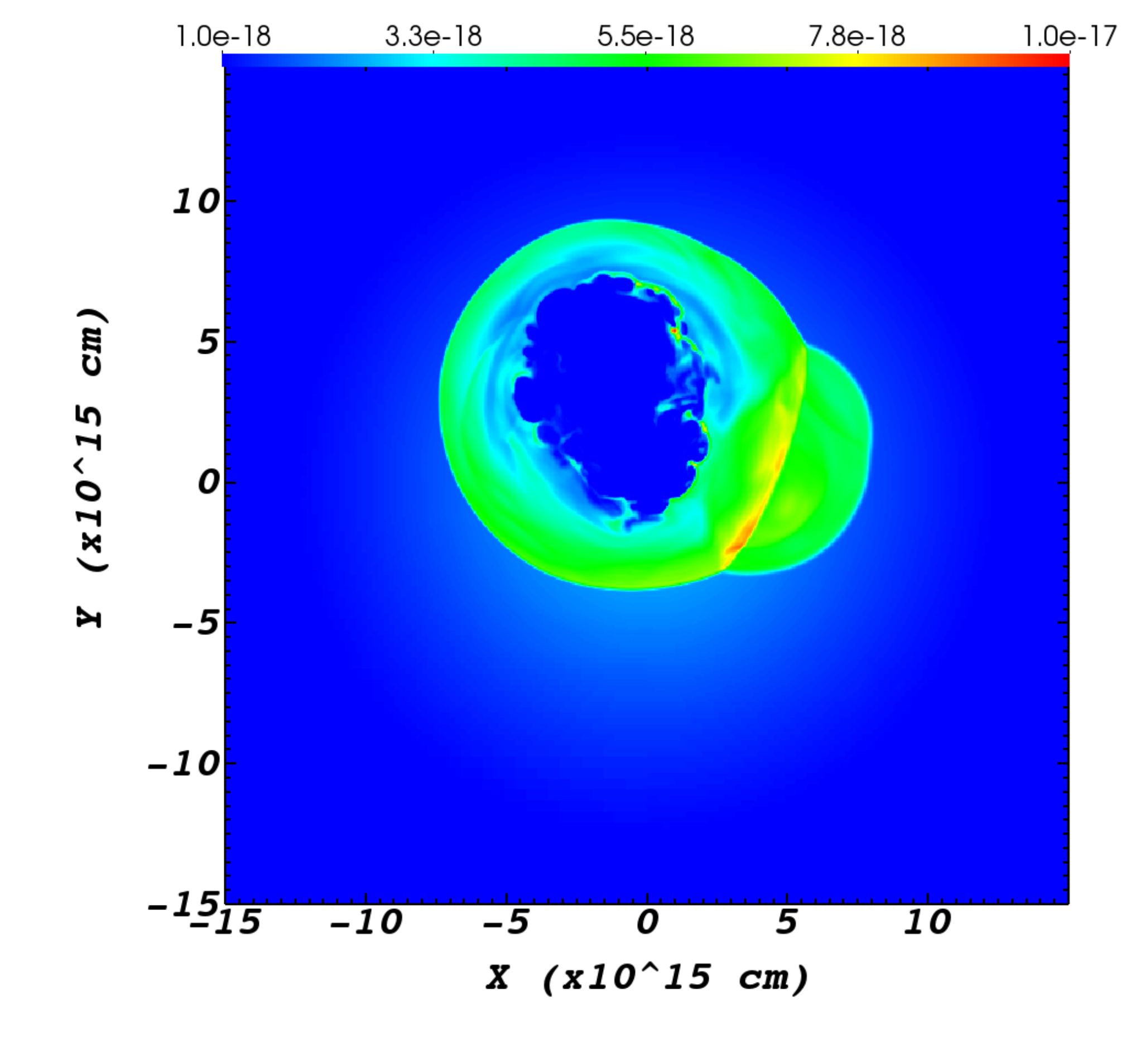}}
\caption{Density maps in planes parallel to the $z=0$ plane at different heights, all at $t=130 \yr$. From upper left, going right and down and in units of $10^{15} \cm$: $z=12$, $z=10$, $z=5$, $z=0$, $z=-5$, and $z=-10$, respectively. Density in $\g \cm^{-3}$ are given by the color bar of each panel. 
}
 \label{fig:cutsxyA}
\end{center}
\end{figure}

\begin{figure}[h!]
\begin{center}
%
%
\subfigure{\includegraphics[height=2.7in,width=2.7in,angle=0]{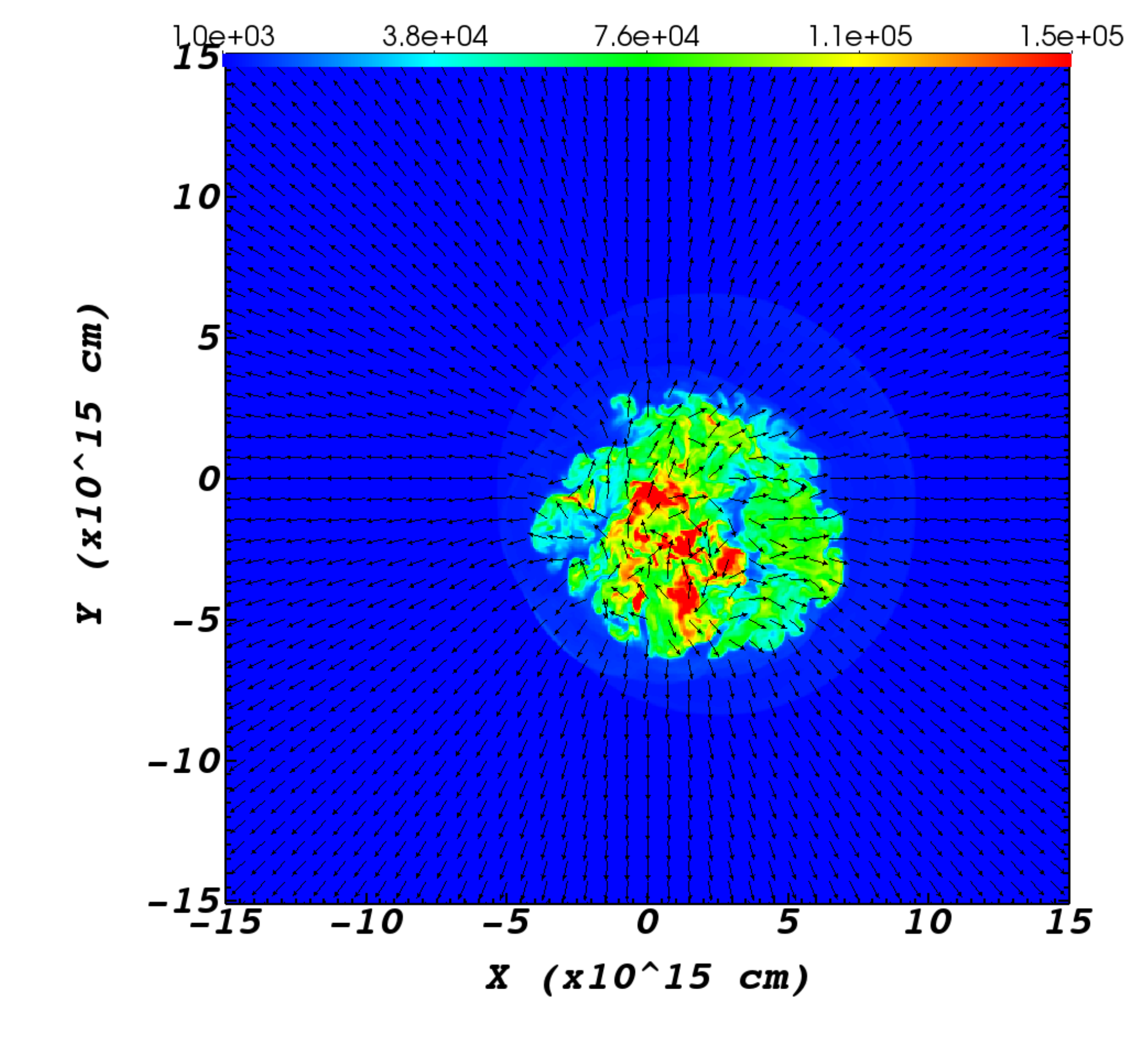}}
\subfigure{\includegraphics[height=2.7in,width=2.7in,angle=0]{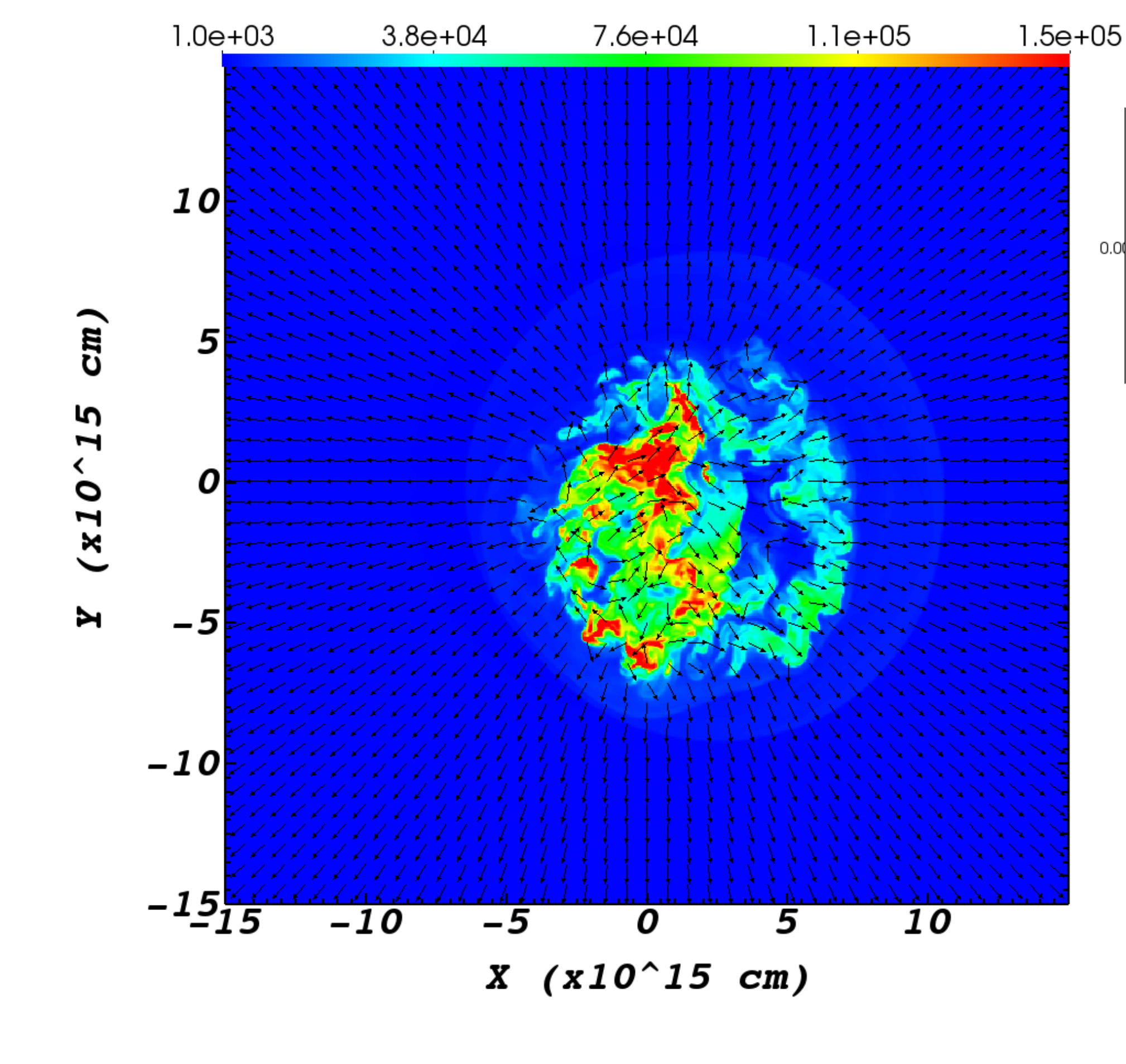}}
\vskip -1. cm
\subfigure{\includegraphics[height=2.7in,width=2.7in,angle=0]{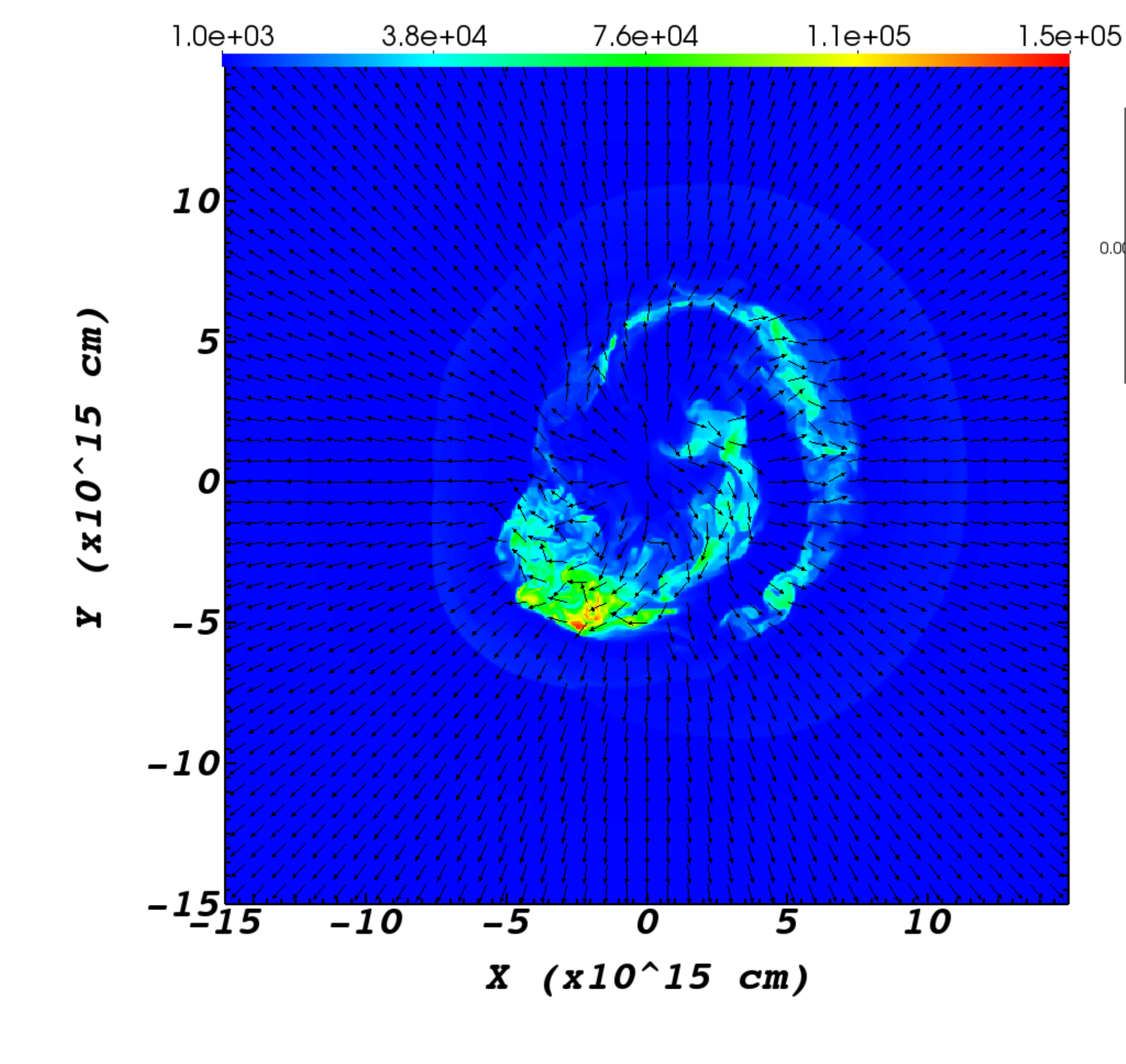}}
\subfigure{\includegraphics[height=2.7in,width=2.7in,angle=0]{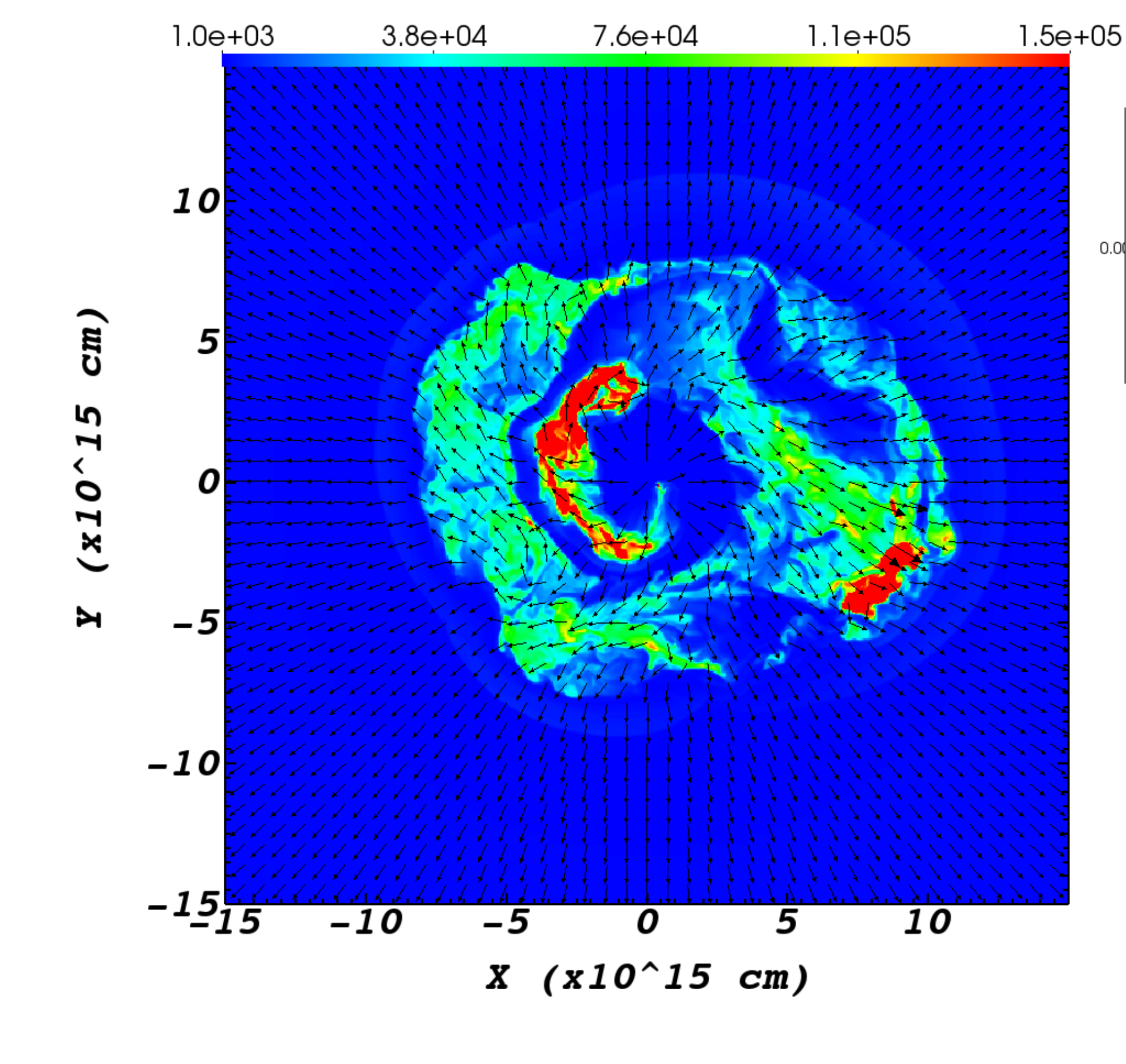}}
\vskip -1. cm
\subfigure{\includegraphics[height=2.7in,width=2.7in,angle=0]{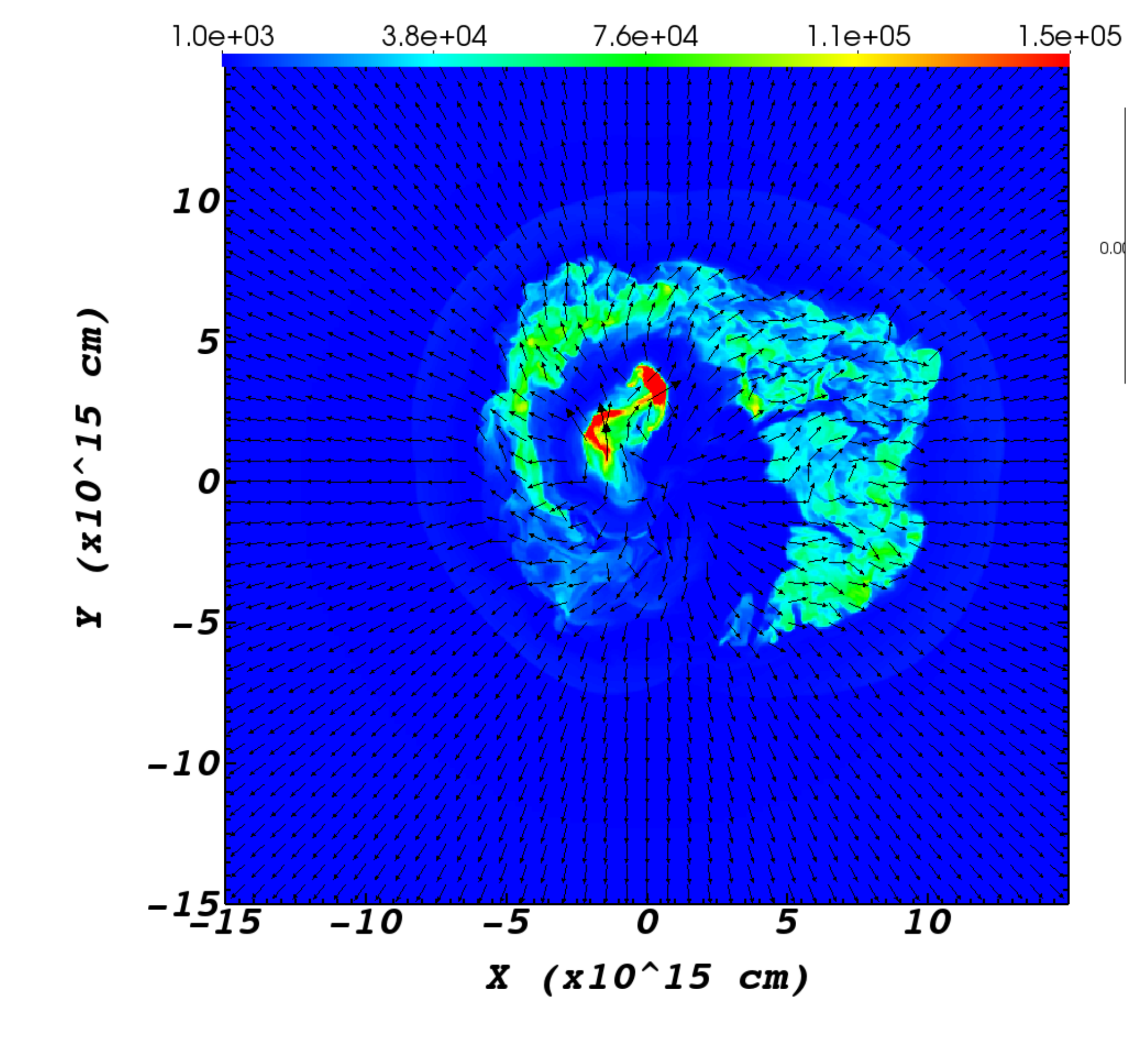}}
\subfigure{\includegraphics[height=2.7in,width=2.7in,angle=0]{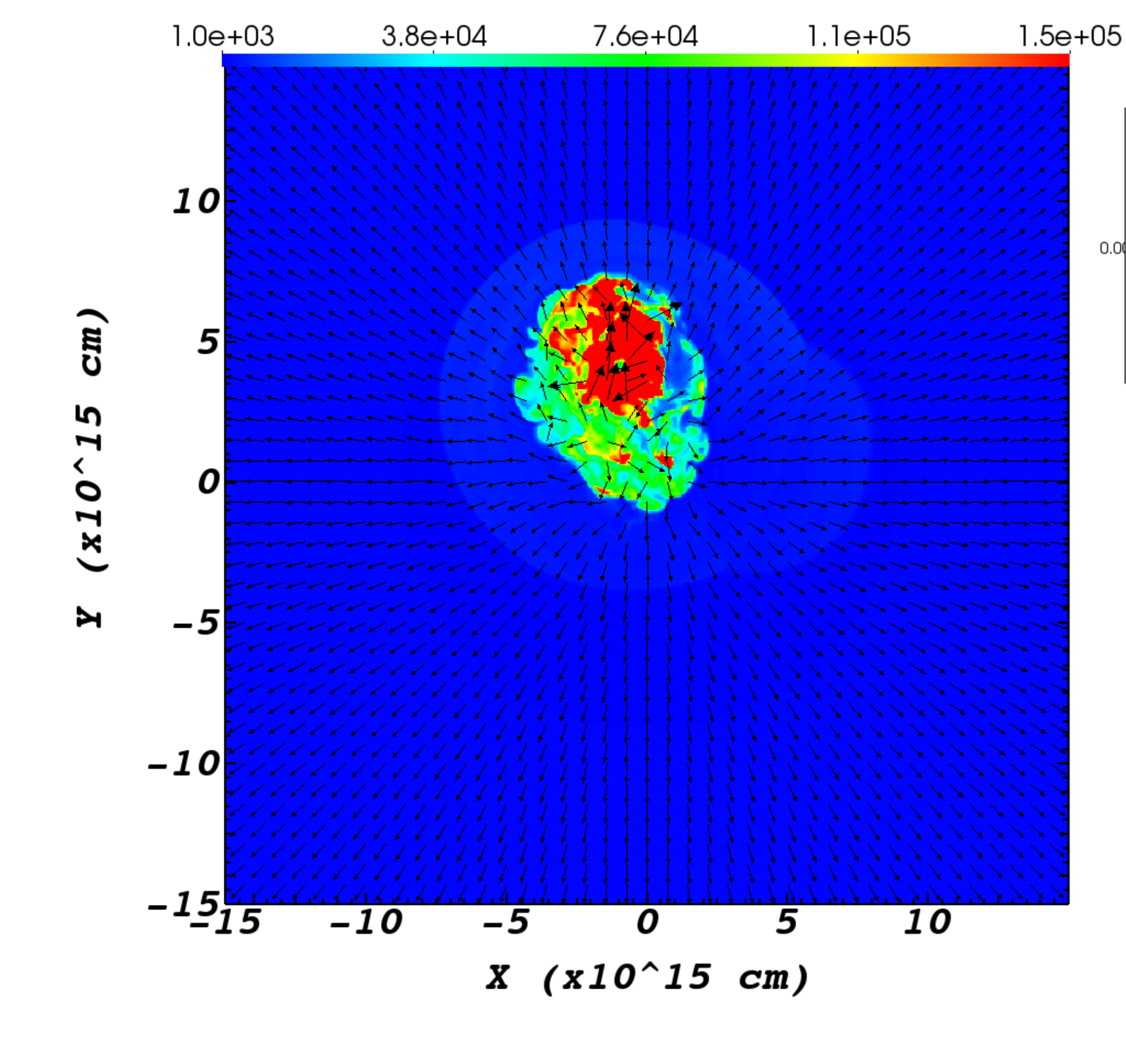}}
\caption{Same as Fig. \ref{fig:cutsxyA} but the gas temperature with the velocity direction.}
 \label{fig:cutsxyB} 
\end{center}
\end{figure}

In Figs. \ref{fig:cutsxyA} and \ref{fig:cutsxyB} we present the density structure and temperature maps with velocity direction, respectively, in six planes parallel to the $z=0$ plane, characterized by their distance from the $z=0$ plane, all at $t=130 \yr$. As before, these panels emphasize the messy nebula and the presence of filaments.  

\begin{figure}[h!]
\begin{center}
%
\subfigure[$(10^{-19}-10^{-18})$]{\includegraphics[height=4.in,width=3.in,angle=0]{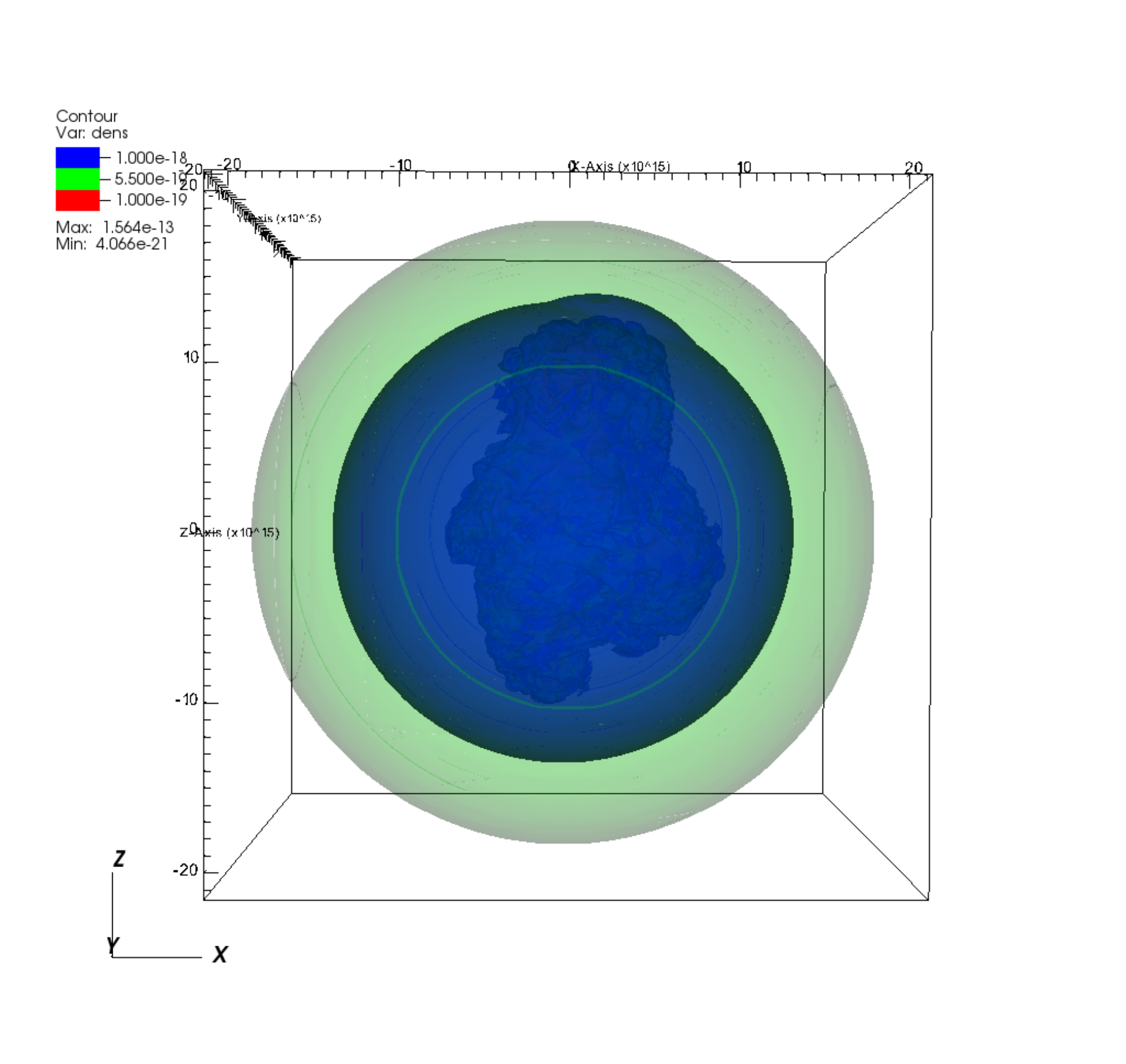}}
\subfigure[$(10^{-18}-10^{-17})$]{\includegraphics[height=4.in,width=3.in,angle=0]{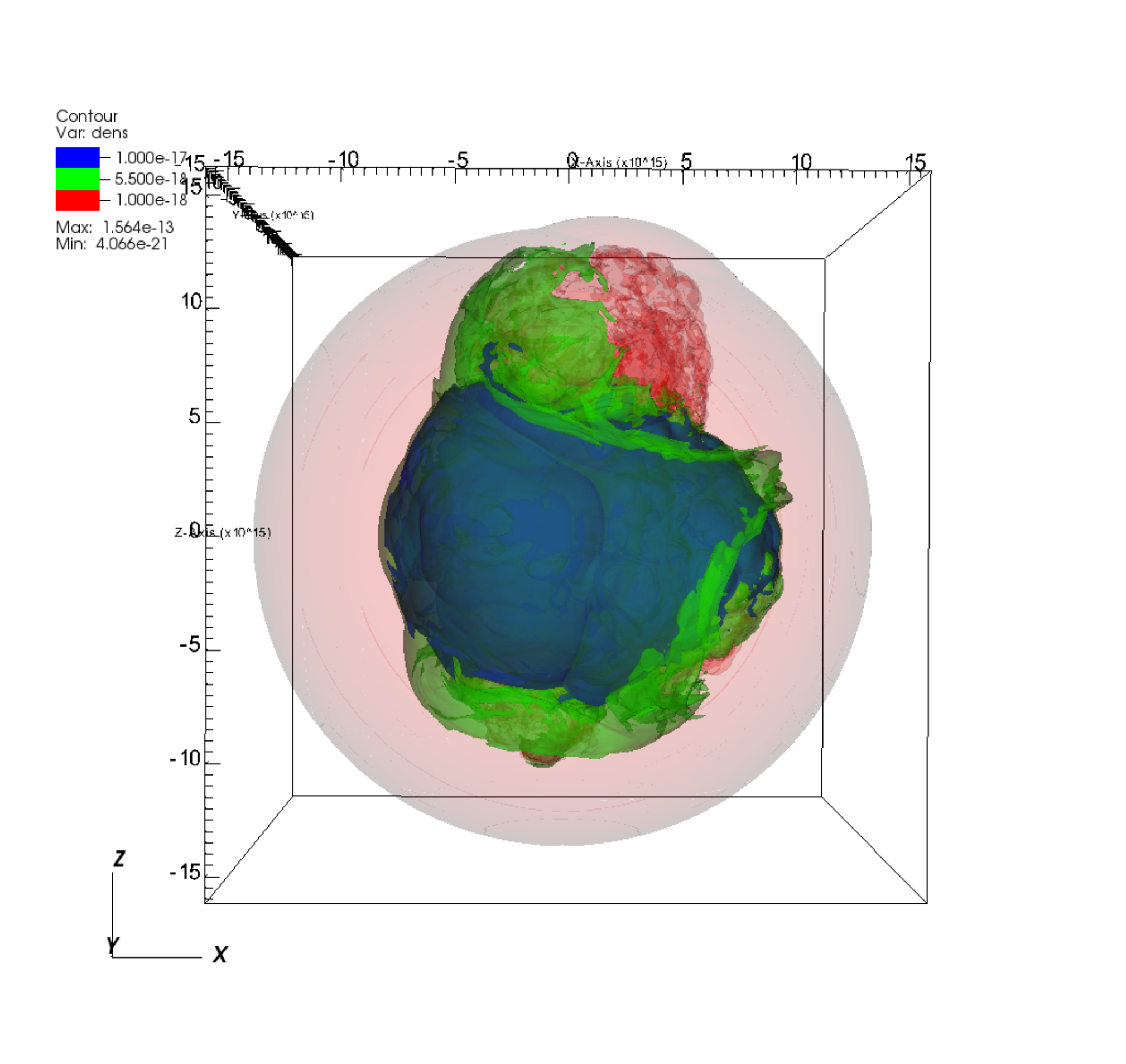}}
\subfigure[$(10^{-18}-5 \times 10^{-17})$]{\includegraphics[height=4.in,width=3.in,angle=0]{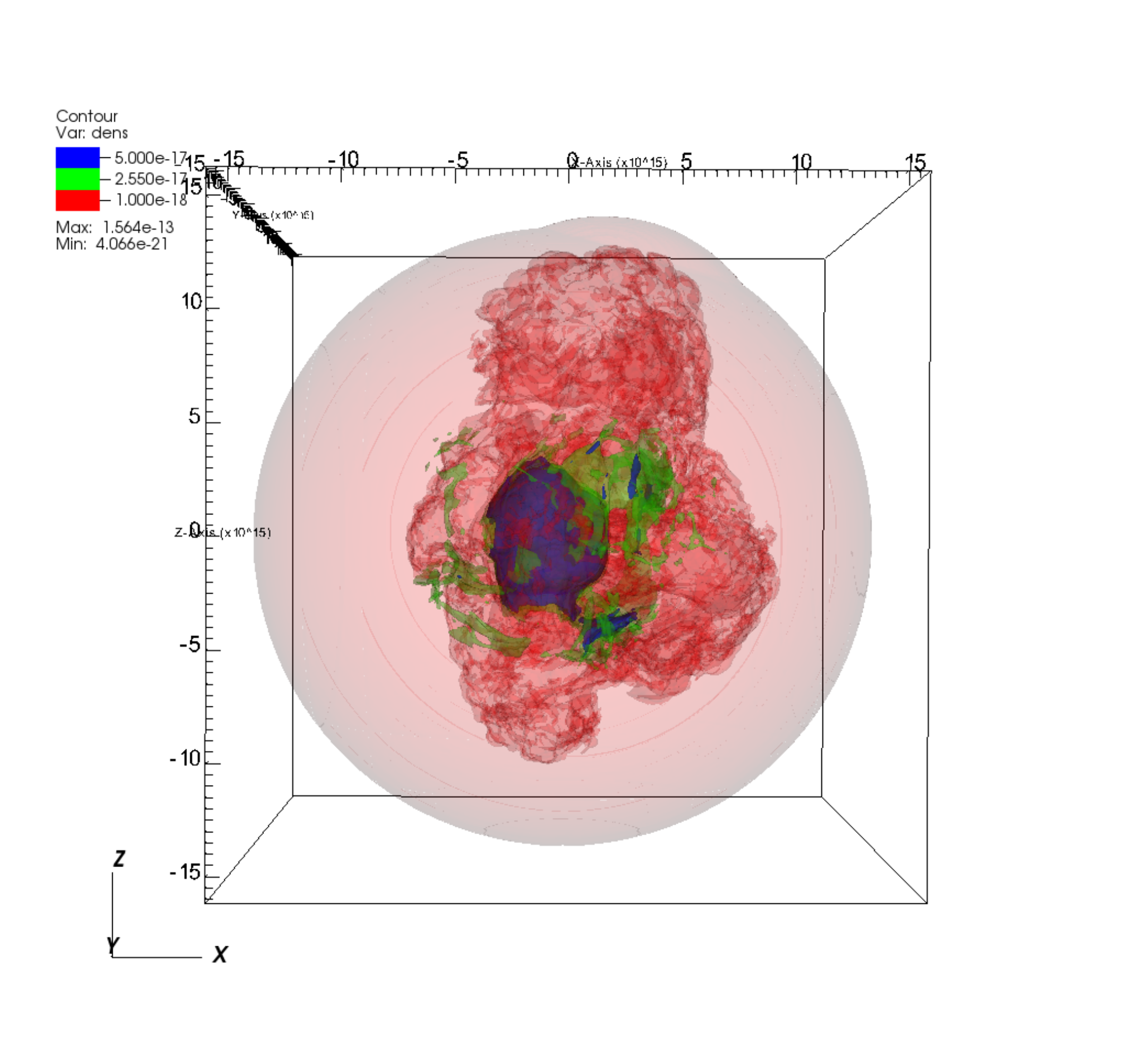}}
\subfigure[$(10^{-17}-10^{-16})$]{\includegraphics[height=4.in,width=3.in,angle=0]{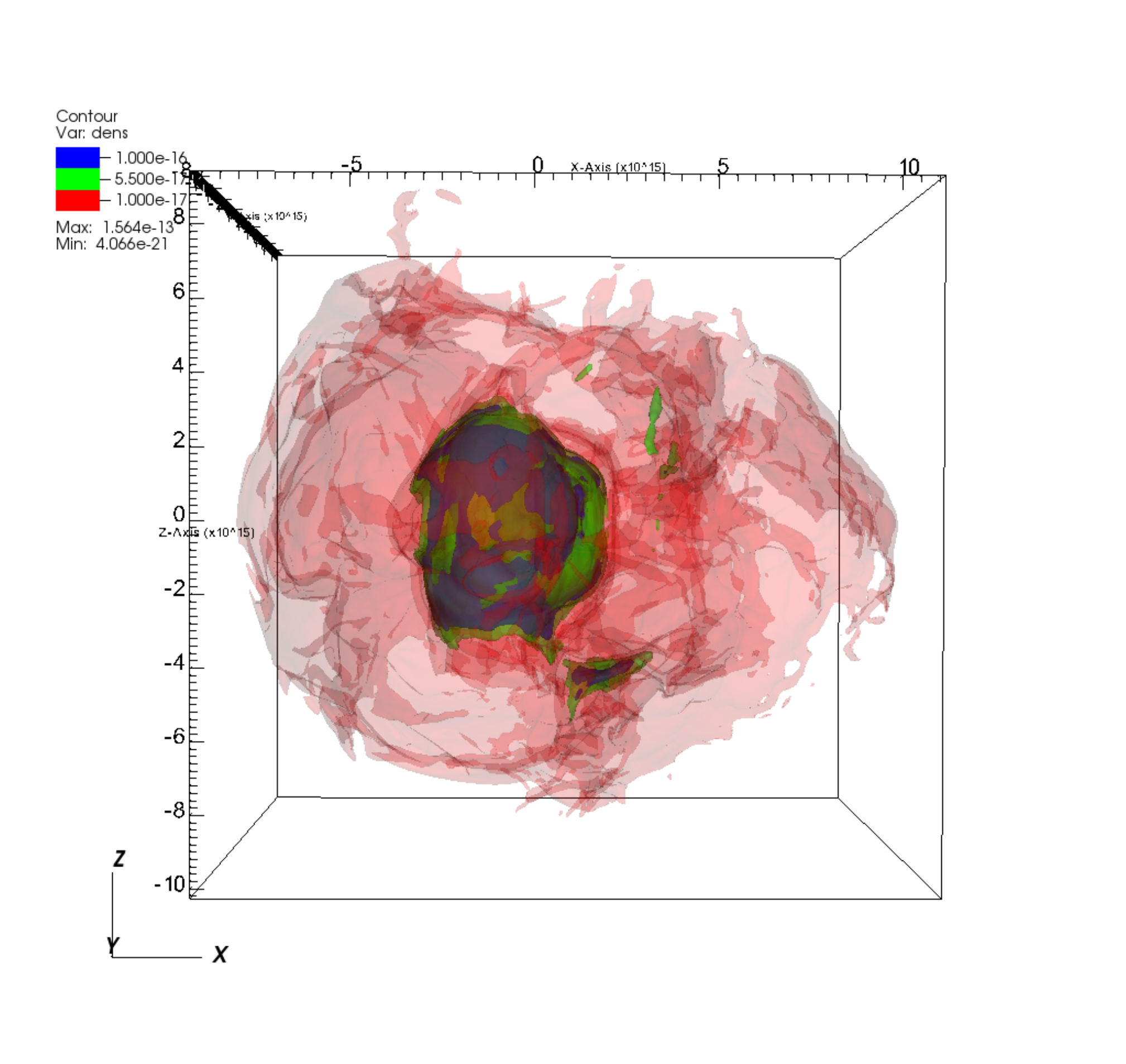}}
\caption{ Three dimensional density structure for the case $\phi = 30$, at $t=130~\yr$.
The range of densities shown by each panel is given in units of $\g \cm^{-3}$.   }
 \label{fig:3D}
\end{center}
\end{figure}

In Fig. \ref{fig:3D}, we present the three-dimensional density structure of the gas at $t=130 \yr$. Each panel emphasizes surfaces with different densities as indicated. These panels fully reveal the messy nature of the nebula and the filaments in the outflow.  

A discussion on the limitations of these simulations is in place here. Ideally we should have run the simulations until the AGB envelope is removed, and then blow the fast wind from the central remnant of the AGB star (the hot core to become a WD). This is impossible due to severe resource limitations when we conduct 3D hydrodynamical simulations. Despite this limitation, for two reasons we argue that our simulations, that show that a messy circumstellar matter is formed, can be projected to the later PN phase. (1) The messy nature of the circumstellar matter that appears after one orbit (panel b in Figs. \ref{fig:rho20degree} and \ref{fig:rho30degree}), does not disappear after 2 orbits (panel d in these figures), and the density contrasts between different segments of the circumstellar matter even increase. So we can project and conclude that running the simulations for over hundreds orbits (thousands of year) till the AGB envelope disappear will not erase the messy nature of the nebula. (2) At the end of our simulations the outer shell expands at a velocity of $v \simeq 50 \km \s^{-1}$, and its sound speed is $C_s \simeq 20 \km \s^{-1}$, i.e., the Mach number is $\mathcal{M} \simeq v/C_s \simeq 2.5$. 
This is high enough to ensure that the messy structure will not be smoothed out.  
We can summarize this point by stating that although we cannot compare this results directly to images of messy PNe, we can confidently claim that the messy circumstellar we have obtained will develop into a messy PN.

\section{SUMMARY}
\label{sec:summary}
 
We conducted 3D hydrodynamical simulations of fast jets interacting with a slow wind from an AGB star. The source of the jets orbits the AGB star with an orbital period of $P_{123}=67~$yrs. 
The significant new ingredient in the simulations was that the symmetry axis of the jets is inclined to the angular momentum axis of the triple-stellar system (see Fig. \ref{fig:schematic}). Such a flow might result if the jets are launched by a star that is a member of tight binary system that orbits the AGB star, accretes mass from its wind, and the orbital plane of the tight binary system is inclined to the orbital plane of the triple system \citep{Soker2004}. 
 
 The evolution of the density structure in one plane is presented in Figs. \ref{fig:rho20degree} and \ref{fig:rho30degree} for two different cases simulated. 
 The density structures in different planes at the end of the simulation for the $\phi=30^ \circ$ case are presented in Figs. \ref{fig:cutsxz} and  \ref{fig:cutsxyA}. These figures clearly show that the nebula lacks any symmetry, and has many filaments. This is termed a `messy nebula'. The high temperature regions that are seen in Figs. \ref {fig:temp30degree} and \ref{fig:cutsxyB} are regions of post-shock jets' material. They have low density, and are termed bubbles. The complicated flow structures of the post-shock jets' material, including vortexes, seen in Figs. \ref{fig:rho20degree}, \ref{fig:cuts_vel}, and \ref{fig:cutsxyB}, account in part for the filamentary structure of the dense gas. 
The full glory of the messy nebula is seen in Fig. \ref{fig:3D}. 
 
One outcome of the motion of the jets in an inclined orbit is that each jet encounters a denser AGB wind than the other jet during half of the orbital period. Over all, the jets do not penetrate in our simulations to large distances, and the bubbles that are formed are small. For example, an `hourglass' nebula is not formed in our simulations. 
One possible example \citep{Soker2004} might be the PN NGC~6210 (images by \citealt{Balick1987, Pottaschetal2009}), that might as well ended in a merger \citep{Soker2016triple}. 

As stated earlier, we cannot follow the evolution to the PN phase, and hence one to one correspondence with observed messy PNe is impossible. However, we can learn from the recent study of \cite{BearSoker2017}. They present images of 4 typical messy PNe, that they claim are most likely have been shaped by a triple stellar system. The two PNe H~2-1 (PN~G350.9+04.4; original image from \citealt{Sahaietal2011}) and Hen~3-1333 (PN~G332.9-09.9 or CPD-568032; original image from \citealt{Chesneauetal2006}) are relevant to our simulations. These two messy PNe lack large lobes and do not have any pair of two opposite lobes. The PN H~2-1 has a bright broken thin shell (a rim), while the PN~Hen~3-1333 has no such rim.  
We expect that the PNe that result from the flow structure we have simulated will look somewhat like one of these messy PNe. Whether a bright thin shell is formed or not, depends on whether the fast wind during the pre-PN phase is strong or not.

We end by crudely estimating the fraction of PNe shaped by the studied scenario. \cite{BearSoker2017} estimate that about $13$ to $21$ per cents of non-spherical PNe have been shaped by triple stellar systems. \cite{Soker2016triple} estimates that about one in eight non-spherical PNe is shaped by the PN AGB progenitor swallowing a binary system or interacting with a binary system close to its surface. 
These are subset of all triple stellar scenarios. The scenario studied in this paper, therefore, accounts for some fraction of the 12 per cents of PNe studied by \cite{BearSoker2017}. We crudely estimate that the scenario studied in this paper accounts for about 3-5 per cents of all non-spherical PNe. There are several scenarios of triple stellar shaping, each accounts for several percents of all non-spherical PNe. The surviving binary system around the central star of the PN is very hard to detect because the binary system is far from the center, and the combined luminosity of the two stars is much smaller than that of the central star. In any case, our results strengthen the call \citep{BearSoker2017} for the search of a triple stellar systems at the center of some messy PNe.
     
This research was supported by the Israel Science Foundation.
N.S. is supported by the Charles Wolfson Academic Chair.


\label{lastpage}

\end{document}